# Recent Advancements in Nanophotonics for Optofluidics

Sen Yang[ae†], Chuchuan Hong[b†], Guodong Zhu[c†], Theodore Anyika[c], Ikjun Hong[c], and Justus C. Ndukaife[cde*]

[a]Institute of Physics, Chinese Academy of Sciences/Beijing National Laboratory for Condensed Matter Physics, Beijing 100190, China; [b]Department of Chemistry and Department of Materials Science and Engineering, Northwestern University, Evanston, Illinois 60208, USA; [c]Department of Electrical and Computer Engineering, Vanderbilt University, Nashville, Tennessee 37235, USA; [d]Department of Mechanical Engineering, Vanderbilt University, Nashville, Tennessee 37235, USA; [e]Interdisciplinary Materials Science Program, Vanderbilt University, Nashville, Tennessee 37240, USA.

*CONTACT Justus Ndukaife. Email: justus.ndukaife@vanderbilt.edu
†These authors contribute equally to this work.

**ABSTRACT**
Optofluidics is dedicated to achieving integrated control of particle and fluid motion, particularly on the micrometer scale, by utilizing light to direct fluid flow and particle motion. The field has seen significant growth recently, driven by the concerted efforts of researchers across various scientific disciplines, notably for its successful applications in biomedical science. In this review, we explore a range of optofluidic architectures developed over the past decade, with a primary focus on mechanisms for precise control of micro and nanoscale biological objects and their applications in sensing. Regarding nanoparticle manipulation, we delve into mechanisms based on optical nanotweezers using nanolocalized light fields and light-based hybrid effects with dramatically improved performance and capabilities. In the context of sensing, we emphasize those works that used optofluidics to aggregate molecules or particles to promote sensing and detection. Additionally, we highlight emerging research directions, encompassing both fundamental principles and practical applications in the field.

## 1. Introduction

*Optofluidics*, as the term suggests, focuses on harnessing the interactions between fluids and light to develop highly adaptable systems. While "fluid" encompasses various states, including gases, our usage specifically refers to liquids. Historically, solid materials have been used in optical systems for millennia, exemplified by the bronze mirrors of China's Shang dynasty (3600 years ago). However, fluids possess unique properties not found in solids, paving the way for innovative optical devices. A primary benefit of fluids is their flowability, enabling changes in the optical characteristics of a fluid medium within a channel simply by swapping fluids or generating gradients in optical properties through mixed fluid flows [1]. The concept of using fluid to control light is not novel; Payne *et al.* introduced the liquid-core fiber optical waveguide in the 1970s [2]. Nonetheless, the field of optofluidics began to take shape only after its successful integration into microfluidic devices, marking the start of an expansive journey guided by the collaborative efforts of the global optical and microfluidic communities. Consequently, optofluidics is defined as "the combination of integrated optical and fluidic components in the same miniaturized system" [3].

Since the 1990s, microfluidic technologies have advanced, enabling precise control of liquids at small scales through intricately designed networks of miniature channels, pumps, valves, and fluid manipulation methods [3]. These devices, with features as small as micrometer scales, are pivotal in various applications, particularly in biotechnology. They offer a wide array of uses, from standard flow cell assays to lab-on-a-chip devices, which integrate comprehensive biochemical synthesis and analysis functions on a compact chip. The advancement of well-developed photonic integration techniques quickly inspired researchers to explore the synergistic combination of micro/nanophotonic devices and microfluidic systems, aiming to enhance both functionality and performance [4–7]. The term *optofluidics* was introduced in 2004 to denote a novel research initiative backed by the Defense Advanced Research Projects Agency (DARPA) of the U.S. Department of Defense [8]. Since its inception, optofluidics has emerged as a prominent research area, witnessing exponential growth in publications and maintaining a high level of activity in the field nowadays.

Optofluidics research takes two primary directions. The first involves creating photonic devices that incorporate fluids, where the optical properties of these fluids are crucial in defining the functions of the devices. The second direction focuses on integrating microfluidic devices with photonic systems, using light to conduct detection and analysis of analytes, or manipulate the fluid as well as colloid particles and ions within these fluids. In the first approach, numerous early experiments have explored various possibilities, including light sources, reconfigurable optical elements, bandpass filters, and more. This diversity stems from the large range of refractive index modulation achievable through fluid manipulation, easily facilitated by microfluidic architectures. Notable examples include Helbo *et al.*'s pioneering chip-scale dye laser in 2003, using a microfluidic channel as a vertical cavity and rhodamine 6G solution for light amplification [9], Chronis *et al.*'s tunable microlens array in 2003, comprising PDMS microfluidic channels filled with liquids and tuned by pressure [10], and Measor *et al.*'s 2010 proposal of an optofluidic bandpass filter employing a liquid-core based anti-resonant reflecting optical waveguide to enhance detection and signal-to-noise ratio [11]. While the rapidly evolving applications of optofluidic devices in biological and chemical sensing have garnered more interest, the exploration of such foundational ideas continues [12]. For instance, Li *et al.* recently demonstrated the comprehensive control of intensity and spectral tuning in all-dielectric metasurfaces through integrated microfluidic channels [13].

Over the past decades, the second research direction, i.e., leveraging optical elements integrated within microfluidic channels for detecting and analyzing target analytes, as well as actively manipulating the fluid or nanoparticles suspended within, has been a focus of intensive investigation in both academic and industrial settings. This interest stems from its immense potential in biological and chemical detection and analysis, and in nanoparticle self-assembly within extremely small volumes (ranging from femtoliters to nanoliters). This integration of sample preparation and delivery with subsequent processing offers substantial benefits, including reduced reagent consumption, portability, cost savings, and increased automation [5]. In this Review, we will stick to this later research direction, *i.e.*, applications of optofluidics in chemical and biological sensing, with a particular emphasis on analytical technologies related to optics and photonics. More importantly, we will delve into various state-of-the-art nanoparticle manipulation techniques. These techniques are crucial for selectively delivering analytes to zones where light-matter interaction is most intense, thereby opening new opportunities in single molecule analysis and single particle spectroscopy to understand the heterogeneity of biological molecules and particles.

As will be illustrated by many of the examples presented in this Review, microfluidics is not just an ancillary component but a crucial integral part of optofluidic devices, significantly expanding the capability to detect analytes. Various optical properties and effects such as refractive index, fluorescence, colorimetry, Raman scattering, and absorption, can be used in optofluidic systems either individually or in combination to produce detectable sensing signals. Recent advancements in nanophotonic structures, including plasmonic nanostructures, photonic crystal cavities, high-Q all-dielectric cavities featuring bound states in the continuum (BICs) [14] and anapole states [15], have enhanced the interactions between target analytes and localized electromagnetic fields. These platforms, compared to traditional integrated photonic structures like optical waveguides, ring resonators, and Fabry–Pérot cavities, offer deeply localized and significantly enhanced electromagnetic fields, thus presenting new opportunities in detecting analytes with low concentrations. For example, changes in the refractive index (RI) of a solution, caused by the presence of analytes, can be identified by observing the optical response of nanophotonic structures. The RI signal typically correlates with the concentration of the analyte or its surface density, rather than the total number of molecules present. Additionally, the intensity of the Surface Enhanced Raman Scattering (SERS) signal tends to be proportional to the fourth power of the local electric field's strength [16]. However, in practice, detection capabilities are often hindered by basic sample delivery systems that fail to selectively and rapidly deliver analytes to zones of strongest light–matter interaction. Common methods to deliver analytes include natural diffusion, flow-over [14], flow-through [17], droplet dropping [18], and spin coating [19]. These methods normally only capture particles close to the optical structure's surface, making them low efficiency for analyte aggregation. Consequently, strategies have been extensively explored to address slow mass-transport issues in optofluidic sensing systems. For example, reports have shown engineered micro/nanoscale geometries passively trapping and separating particles using elastomeric collapse techniques [20,21]. Mechanisms including particle transport and trapping by optical forces or non-optical forces [22–27] are also actively investigated. It should be noted that achieving continuous mass transport to steadily enhance the analyte concentration at the detection region is desired for high-sensitivity detection. This is because it alters the equilibrium of the concentration distribution of the analytes, using an external energy to overcome the diffusion limit [28]. Moreover, the capability to rapidly transport and stably hold a single particle at the detection site is particularly important, as it enables the study of individual target particles or molecules to understand their heterogeneity rather than those of a group.

This review presents a brief overview of recent progress in optofluidic applications, particularly highlighting the state-of-the-art particle manipulation techniques in optofluidic channels, and recently developed optofluidic systems for biological and chemical sensing applications. First, we explore particle control mechanisms, dividing them into two categories: those involving solely optical forces and those incorporating hybrid effects in microfluidic channels, such as electro-osmotic and electrothermoplasmonic flows. Subsequently, we examine four different sensing and detection mechanisms that integrate optical nanostructures in microfluidic channels. Finally, we will discuss the outlook of this field, addressing various challenges and opportunities both from a fundamental and an application perspective.

## 2. Nanoparticle manipulation in microfluidic chambers

The flexible and precise control of particles, encompassing their transport, trapping, and manipulation (such as movement and rotation), is crucial for the detection and analysis of target analytes in optofluidic systems. These control techniques can be categorized into two distinct groups: those governed solely by optical forces and those driven by photo-induced effects. We will discuss each of these categories and highlight latest achievements in the field in the subsequent sections.

### 2.1 Optical trapping

Since Arthur Ashkin's pioneering demonstration of trapping dielectric particles by an optical tweezer using a tightly focused laser beam in 1986 [29], this technology has become an indispensable tool in the life sciences for manipulating microscale objects, including bacteria, cells and even DNA molecules tethered to micron scale beads. Arthur Ashkin's groundbreaking work in this field was honored with the Nobel Prize in Physics in 2018 [30]. A typical single-beam optical tweezer employs a tightly focused laser beam to trap particles near its focal point. The particle, once stably trapped, is subject to two primary forces [31,32]: the optical gradient force $\mathbf{F_{grad}}$, as defined in Equation (1), and the optical scattering force $\mathbf{F_{scat}}$, as defined in Equation (2). Notably, the optical gradient force is pivotal for trapping, as it directs the particle towards the focal spot and must overcome the scattering force for trapping to occur. The expressions for these forces in the quasi-static limit are provided in the following.

$$\mathbf{F_{grad}}(\mathbf{r}) = \pi \varepsilon_e a^3 \frac{\varepsilon_p - \varepsilon_e}{\varepsilon_p + 2\varepsilon_e} \nabla |\mathbf{E}(\mathbf{r})|^2 \quad (1)$$

$$\mathbf{F_{scat}}(\mathbf{r}) = \frac{128\pi^5 a^6}{3\lambda^4 c} \left(\frac{\varepsilon_p - \varepsilon_e}{\varepsilon_p + 2\varepsilon_e}\right)^2 \mathbf{S}(\mathbf{r}) \quad (2)$$

Here $\varepsilon_p$ and $\varepsilon_e$ are the permittivity of the particle and the environment, respectively; $a$ is the radius of the particle; $\mathbf{S}(\mathbf{r})$ is the time-averaged Poynting vector defined as $\frac{1}{2}\mathbf{Re}\left[\mathbf{E} \times \mathbf{H}^*\right]$ where **E** and **H** are local electric field and magnetic field, respectively.

Optical tweezers offer flexible three-dimensional control but integrating them with microfluidic systems is challenging due to the external light field requirement. To incorporate particle manipulation capabilities onto microfluidic chips, early efforts utilized mature integrated photonic elements like optical waveguides and microring resonators. For instance, Lin *et al.* demonstrated the effective optical manipulation of micrometer-sized dielectric particles using a planar silicon microring resonator [33]. Here, particles in a fluid above a bus optical waveguide are attracted to the optical waveguide surface by the optical gradient force and then propelled along the direction of energy flow by the optical scattering force. The velocity is determined by the balance between optical and viscous drag forces. As a particle reaches the microring resonator's top surface, it orbits around the ring at several hertz. Despite these remarkable achievements, the size of particles that can be trapped by optical tweezers and the evanescent fields from such optical waveguide systems is constrained by the diffraction limit of light [34], which hinders the focusing of light to nanoscale subwavelength volumes. To trap smaller particles, particularly those under 100 nm that are of significant interest in life sciences, such as protein molecules, extracellular vesicles (EVs) and non-vesicular extracellular nanoparticles [35], the development of optical nanotweezers has been pivotal.

### 2.1.1 Plasmonic nanoantenna tweezers

Plasmonic nanoantennas have the capability to confine light to deeply subwavelength volumes to create high field enhancements at the nanoscale. These subwavelength hotspots can generate the tightly confined trapping potential wells necessary for trapping nanoscale objects. Following early works by Novotny, Quidant and others on plasmonic forces [36,37], plasmonic optical tweezers capable of trapping micrometer-scale dielectric beads were experimentally reported by Righini *et al.* [38,39] (see Figure 1(a)) and the size of trapped beads was pushed down to 200 nm by Grigorenko *et al.* [40]. By utilizing electromagnetically coupled pairs of gold nanodots, Grigorenko *et al.* significantly reduced the trapping volume to subwavelength scales. They observed an almost tenfold increase in particle confinement compared to traditional optical tweezers. These early promising research results spurred intense research in the field, leading to the development of various types of plasmonic nanotweezers, including nanoantenna structures like nanopillars/disks [31], dimers (Figure 1(b)) [41], bipyramidal nanoantennas [42], bowties (Figure 1(c)) [43], and nanoaperture structures such as nanoholes [44], double nanoholes [45], coaxial apertures (Figure 1(d)) [46], and bowtie apertures (Figure 1(e)) [47,48]. Each of these plasmonic nanotweezers operates on the principle of localized surface plasmon resonance, with different nanoantenna designs yielding distinct near-field distributions to create highly localized and deep trapping potentials. For instance, Au nanoparticles as small as 10 nm were successfully trapped in the gap of plasmonic dipole antennas [41], as depicted in Figure 1(b). Wang *et al.* were the first to report the rotation of dielectric nanoparticles as small as 110 nm in diameter trapped around a pillar-like plasmonic nanoantenna [49]. This rotation was achieved either by manually tuning the polarization of the linearly polarized trapping laser beam or by using a circularly polarized illumination.

While nanoscale trapping using plasmonic nanoantenna structures has been successfully demonstrated, a significant challenge arises from the substantial thermal gradients at the hotspots due to the inherent lossy nature of these antennas [43,50]. This phenomenon results in positive thermophoretic forces and Rayleigh-Bénard fluid forces, surpassing the optical gradient force and leading to the repulsion of particles from the hotspot. Plasmonic nanoapertures have been demonstrated to effectively trap nanoscopic particles, achieving success even with particles in the sub-10 nm regime. This efficacy is attributed to the concentrated intensity of light in the gap region of the structure. Notably, apertures in metallic films provide more efficient dissipation of heat away from the hotspots due to the high thermal conductivity of the extended metallic film, thereby reducing the impact of thermal effects and facilitating the trapping of very small objects. Based on gold nanohole structures, Quidant's group was the first to demonstrate self-induced back-action (SIBA) [44]. In the particle−cavity system, the trapped particle alters the local electric field, actively influencing the trapping dynamics by pulling itself back towards the cavity when it strays from its equilibrium position. By harnessing the SIBA mechanism, the power required for stable trapping is reduced, permitting low power nanoscale optical trapping. Similarly, they also showed that a bowtie-shaped plasmonic nanoaperture created on a gold film at the end of a tapered metal-coated optical fiber achieved three-dimensional optical manipulation of single 50 nm dielectric nanoparticles by leveraging the SIBA effect [47]. This non-invasive method is promising for the flexible control of nanosized objects. Plasmonic double nanoholes (DNH), as another aperture-based nanotweezer, have also been widely explored. Besides trapping very tiny nanoparticles (sub-20 nm) [45,51], they have also been used for detecting biomolecules and analyzing single-molecule kinetics [52–54]. However, traditional DNH apertures which are typically fabricated directly on a glass substrate face a common challenge—they are usually illuminated off-resonance. Operating with resonance detuning results in less intense electromagnetic hotspots. This implies that these systems are not fully capitalizing on the benefits of the plasmonic aperture to maximize the optical gradient force and to enhance the spectroscopy of trapped specimens. Recently, Anyika *et al.* [55] have shown that by introducing a reflector layer sitting on a sapphire substrate to the DNH structure (as depicted in Figure 1(f)), the plasmon resonance can couple the reflected light, establishing the metallic film thickness as an additional degree of freedom for resonance tuning, without the requirement of shrinking the size of the aperture. Such a design enables on-resonance illumination of DNH apertures while efficiently dissipating heat in both in-plane and axial directions, leading to minimal temperature rise. Using this approach, low-power trapping of small extracellular vesicles have been demonstrated.

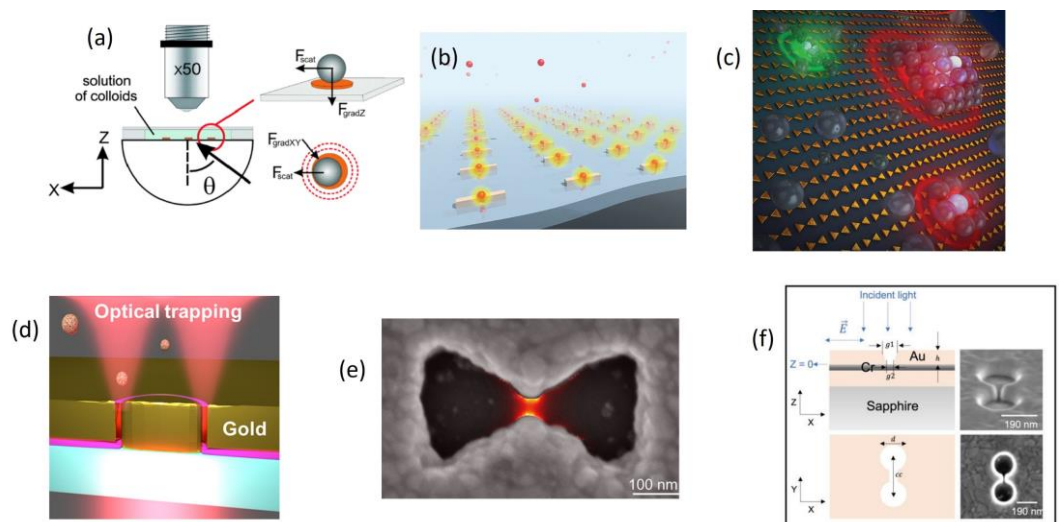

**Figure 1** Typical plasmonic nanotweezers. (a) describes the set-up of the plasmonic nanotweezers composed of isolated gold microdisks fabricated on a glass substrate. A 4.88 μm polystyrene (PS) microbead is trapped on a microdisk. (a) adapted with permission from ref [39]. Copyright 2008 American Physical Society. (b) shows trapping of 10 nm Au nanoparticles with arrays of plasmonic dipole antennas. (b) adapted with permission from ref [41]. Copyright 2010 American Chemical Society. (c) shows the use of Au bowtie nanoantenna arrays for optical trapping, stacking, and sorting. (c) adapted with permission from ref [43]. Copyright 2012 American Chemical Society. (d) demonstrates the optical trapping of 30 nm polystyrene beads and streptavidin molecules with a 10 nm gap resonant coaxial nanoaperture in a gold film. (d) adapted with permission from ref [46]. Copyright 2018 American Chemical Society. (e) presents the optical trapping of individual silica-coated quantum dots by bowtie apertures fabricated in a silver film. (e) adapted with permission from ref [48]. Copyright 2016 American Chemical Society. (f) depicts the cross-section schematics and scanning electron microscope (SEM) images for a DNH with a reflector layer. The structure was illuminated under *x* polarized light. (f) adapted with permission from ref [55]. Copyright 2023 American Chemical Society.

### 2.1.2 Dielectric nanoantenna, photonic crystal and metasurface tweezers

Numerous strategies have been proposed to minimize heating side effects in optical trapping, which can destabilize the trap due to thermophoresis and convection or pose damage to delicate biological specimens. However, these issues are often inevitable due to the inherent lossy nature of metallic materials used in plasmonic approaches. To address these challenges, non-plasmonic methods, *i.e.,* **dielectric nanotweezers** are considered as viable alternatives and have generated significant attention in recent years. Overcoming the fundamental diffraction problem in optical tweezers requires strong localization of the electric field, a task not easily achieved with traditional integrated photonic devices like optical waveguides. Yang *et al.* tackled this challenge by incorporating a sub-wavelength slot into waveguides [56].

The significant electric field discontinuity at the horizontal boundaries of the slot region results in the electric field being intensely localized within the gap. This innovation enables stable trapping and transport of nanoparticles and biomolecules as small as 75 nm. The team also numerically showed in a later work that such stable trapping and transport can be achieved for objects as small as 10 or 20 nm in diameter, as illustrated in Figure 2(a) [57].

The localization of the electric field in photonic crystal (PhC) cavities provides another solution. Both one-dimensional [27,58,59] (Figure 2(b)) and two-dimensional PhC cavities [60,61] (Figure 2(c)) with a point defect at the center of the structure have been reported. SIBA was experimentally confirmed in a two-dimensional PhC cavity by Descharmes *et al.* [60], validating its universality in particle-cavity systems during trapping processes. Additionally, these systems permit highly sensitive detection of trapped specimens [62,63]. This advantage stems not only from the naturally high Q factors characteristic of PhC resonators but also from the exposed electric field in the air holes of the PhCs, which facilitates significantly stronger light-matter interactions in comparison to conventional slotted waveguides. Mandal *et al.* utilized a one-dimensional photonic crystal resonator array to achieve device sensitivity an order of magnitude higher than that of whispering gallery mode type biosensors [62]. Moreover, Lin *et al.* reported SERS detection in a reproducible manner by trapping analyte-conjugated Ag particles [64]. This was made possible as the optical trapping process could be monitored in real-time through the cavity resonance shift occurring with the trapping of each additional nanoparticle, showcasing the potential of PhC cavities in advanced sensing applications.

Recent advancements in meta-optics and nanophotonics driven by Mie resonances [65], have opened new avenues for designing dielectric nanotweezers. These developments offer some advantages over PhC-based nanotweezers: the size of isolated Mie-based antennas is substantially smaller, resulting in a more compact footprint. This allows for multiple trapping sites when arranged in an array. In the case of Mie-based metasurfaces, the interaction area between analytes and the localized electric field is expanded, courtesy of periodically arranged hotspots. Furthermore, these systems generally utilize normal incidence illumination for excitation, which is highly compatible with standard microscope systems and differs markedly from the waveguide and/or optical fiber coupling used in PhC systems. However, the exploration of particle manipulation and sensing applications in these systems is still in its nascent stages compared to the more established field of plasmonic nanotweezers. This area is ripe with exciting opportunities awaiting further exploration and requires dedicated effort.

In this context, we present three examples that illustrate the initial investigations in this promising field. Xu *et al.* successfully demonstrated the optical trapping of polystyrene spheres [66] and quantum dots [67] using an all-dielectric system, as shown in Figure 2(d). In this setup, particles are trapped at the gap between two identical silicon cylinders. The authors noted that while the electric field enhancements in this system are modest (only 5 times enhancement) compared to plasmonic nanotweezers, they are still sufficient to trap relatively high-index nanoparticles (NSs) as small as 20 nm, with minimal heat generation (a temperature rise below 0.04 K, in contrast to tens of Kelvin with gold resonators). To optimize the optical gradient force on a nanoscale particle and reduce the required laser power for optical trapping, enhancing and exposing the electric field from the dielectric antenna is the key. The optical anapole state [15] characterized by the destructive interference of the electric and toroidal dipole radiations in the far field, thus enhancing the electromagnetic field in the near field, has emerged as a promising approach. This approach significantly increases the optical gradient force on a nanoscale particle, leading to effective and stable optical trapping. Hernández-Sarria et al. theoretically showed that a single nanodisk resonator with a rectangular-shaped hole supporting the anapole state can create a strong electromagnetic hotspot, capable of trapping a sub-20 nm nanoparticle using an electric field enhanced by the anapole state over 10 times [68]. They later further presented the capability of such a system to trap one or two nanoparticles with different morphologies [69]. Hong *et al.* successfully demonstrated the use of optical anapole states for low-power stable trapping of nanoscale extracellular vesicles and the recently discovered non-vesicular extracellular nanoparticles known as supermeres [70]] through experiments, as shown in Figure 2(e). This was achieved by utilizing a distributed Bragg-reflector layer as a substrate in conjunction with a double nanohole aperture. This innovative approach makes the electromagnetic field accessible to nanoscale particles while effectively minimizing heating effects. Conteduca *et al.* theoretically predicted [71] and experimentally demonstrated [72] the parallel trapping of tens of 100 nm polystyrene particles with a laser power density as low as ∼ 160μW/μm² by a dielectric metasurface composed of periodically arranged nanocuboid antennas supporting the anapole states.

Dielectric metasurfaces can also achieve efficient optical trapping. Recently, the concept of bound states in the continuum (BICs) [73], which were originally proposed in quantum mechanics [74] and refers to a kind of nonradiating state of light in photonics, has garnered significant interest. This is due to its potential to enable metasurfaces to achieve high field enhancement comparable to or even higher than that in plasmonic systems, and Q factors comparable to PhC cavities [19,75]. This makes BICs an attractive option for trapping-assisted sensing applications. Yang *et al.* reported the first theoretical demonstration of stable optical trapping of sub-10-nm nanoparticles using low laser power in a quasi-BIC system [76], as illustrated in Figure 2(f). Their findings also revealed that trapped particles could enhance, rather than suppress, the resonance mode of the cavity when the system's mirror symmetry is broken. This enhancement of the trapping process introduces a novel mechanism for inducing SIBA. Moreover, Conteduca *et al.* have very recently achieved the experimental realization of optical trapping using a similar quasi-BIC metasurface, marking an important advancement in this area [77].

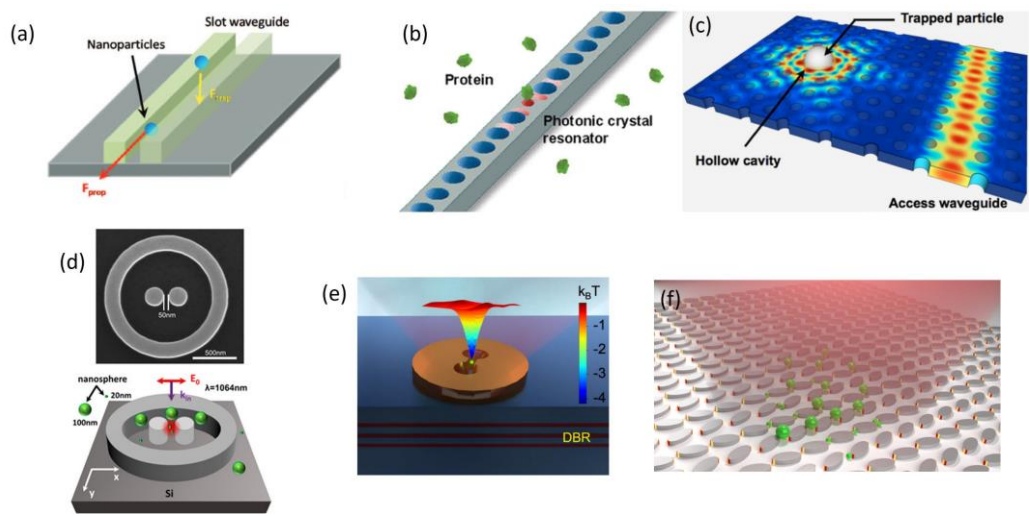

**Figure 2** Typical dielectric nanotweezers. (a) presents the forces acting on nanoparticles in a slot waveguide made of silicon immersed in water. Nanoparticles will be trapped towards the slot where the high-intensity slot mode is located. (a) adapted with permission from ref [57]. Copyright 2009 American Chemical Society. (b) shows the trapping of a single protein molecule on a one-dimensional silicon nitride PhC resonator. (b) adapted with permission from ref [58]. Copyright 2012 American Chemical Society. (c) presents the electric field distribution in a two-dimensional PhC cavity in the presence of a 500 nm dielectric particle. (c) adapted with permission from ref [60]. Copyright 2012 American Physical Society. (d) depicts the SEM image and schematic illustration of the optical trapping of nanospheres using an all-Si nanoantenna. Particles as small as 20 nm in diameter were optically trapped. (d) adapted with permission from ref [66]. Copyright 2018 American Chemical Society. (e) describes the trapping potential when trapping a single nanoparticle in the double-nano slot of an anapole nanoantenna. The electromagnetic field is enhanced aided by the bottom Bragg reflector layer. (e) adapted with permission from ref [70]. Copyright 2023 American Physical Society. (f) shows parallel trapping of multiple nanoparticles on a quasi-BIC metasurface. Gradient colors at the tips depict the electric field distribution. (f) adapted with permission from ref [76]. Copyright 2021 American Chemical Society.

### 2.1.3 Metasurface lens (metalens) tweezers

Microlens and metalens [78] present an intriguing alternative for miniaturized optical tweezers. Despite the diffraction limit constraining the smallest size of particles that can be trapped, the extensive tunability of metalens designs enables a wide range of particle trapping controls. For instance, Markovich *et al.* demonstrated the ability of a bifocal Fresnel metalens to create two stable trapping centers along the optical axis [79], as depicted in Figure 3(a), with the trapped particle's position controlled by the polarization of the incident light. Chantakit *et al.* achieved similar control over focal spots along the axial direction using a polarization-sensitive Pancharatnam–Berry phase bifocal metalens [80], as shown in Figure 3(b). They showcased drag-and-drop manipulation of polystyrene particles and demonstrated the transfer of angular orbital momentum to these particles using a single tailored beam. Ma *et al.* theoretically suggested that high- and low-refractive-index (index higher and lower than the liquid media, respectively) particles could be simultaneously trapped by a multifocal optical vortex metalens [81]. Zhu *et al.*, through numerical calculations, indicated that longitudinal and transverse optical trapping could be achieved by toggling between focusing radial and azimuthal vector beams with a dual-focusing metalens [82]. Microlenses and metalenses can also be integrated with other systems for multifunctional particle manipulation in a compact setup. Li *et al.* combined microlenses with an optical fiber probe, enabling both three-dimensional manipulation and real-time high-sensitivity detection [83], as illustrated in Figure 3(c). Wang *et al.* showed that combining polarization-sensitive metalens with a plasmonic film could facilitate selective trapping of metallic nanoparticles [84]. By leveraging the effects of two orthogonal circular polarizations, single particles can be stably trapped at the center while repelling all others. Additionally, Hsu *et al.* reported the possibility of trapping and imaging single atoms by combining a magneto-optical trap with a polarization-gradient cooling setup [85]. Such achievement may enable advanced control in complex quantum information experiments.

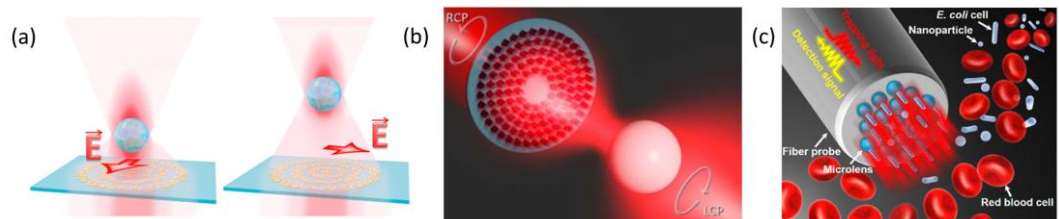

**Figure 3** Typical optical tweezers based on microlens and metalens. (a) illustrates the control of an optically trapped particle's position along the optical axis by altering the polarization of the incident laser beam that illuminates the metalens. (a) adapted with permission from ref [79]. Copyright 2018 American Physical Society. (b) depicts the trapping of a polystyrene microbead using an all-dielectric metalens, which transforms a right circularly polarized incident Gaussian beam into a focused left circularly polarized beam. (b) adapted with permission from ref [80]. Copyright 2020 Chinese Laser Press. (c) shows the schematic illustration of selective trapping and detection of multiple nanoparticles by a photonic nanojet array assembled on an optical fiber probe. (c) adapted with permission from ref [83]. Copyright 2016 American Physical Society.

## 2.2 Hybrid optically-induced approaches

In our prior discussions, we have highlighted how optical forces have opened numerous possibilities in nanoparticle manipulation. The advent of optical tweezers brought about the flexible manipulation of large biomaterials, such as cells. This was further advanced by the introduction of plasmonic and dielectric nanotweezers, which extended the trapping capabilities to much smaller particles, including proteins and viruses. Nonetheless, there are inherent limitations, such as the diffraction limit of light [75] and the potential for photoinduced damage to biological particles [76] in optical tweezers, as well as challenges like the slow rate of transporting and delivering target particles to the electromagnetic hot spots, and the lack of dynamic manipulation in both plasmonic and dielectric antenna nanotweezers [26]. Overcoming these obstacles is imperative. Recent advancements show the integration of nanophotonics with other mechanisms to develop hybrid nanomanipulation approaches, offering fresh perspectives and opportunities to enhance the field of light-based nanomanipulations.

Photo-induced heating due to Ohmic loss is an inevitable topic in the applications of plasmonic materials. In plasmonic nanotweezers, heat generation also needs to be managed carefully [86–92]. On the one hand, excessive temperature increases due to photo-induced heating are often viewed as detrimental to trapping events [90,93]. Such heat can cause thermal damage to delicate biological specimens [94,95], and give rise to various fluidic phenomena that may disrupt the optical trapping process [93,96]. On the other hand, the temperature rise can be reduced or even eliminated after introducing rapid heat dissipation approaches, such as using high thermal-conductivity substrates [55,97–99]. However, photo-induced heating can also be strategically leveraged to enhance the trapping stability, to facilitate rapid particle transfer, or to enable particle aggregation and assemblies. Therefore, the following sections of this chapter will review the latest developments in light-based hybrid nanoparticle manipulation methods, mainly focused on heat-assisted and electrohydrodynamic-assisted nanomanipulations.

### 2.2.1 Electrothermoplasmonic tweezers

**Electrothermoplasmonic (ETP)** flow harnesses photo-induced heating and applied AC electric field [26,100,101] to rapidly load a near-field plasmonic trap and to get access to the nanoparticles hundreds of micrometers away from the plasmonic nanocavity. It exemplifies the advantages of hybrid optically-induced nanomanipulation approaches.

In general, the generation of ETP flow can be explained as follows [26,102]. In water, when the temperature gradient is generated by photo-induced heating upon laser illumination of a plasmonic cavity, the local fluid permittivity and electric conductivity are changed spatially since they depend on the local temperature. Upon applying an external AC electric field, the AC field acts on these gradients in the permittivity and electric conductivity, generating a body force on the fluid to drag and transport particles in the solution. The time-averaged electrical body force per unit volume at frequency ω can be expressed as:

$$\langle F_{ETP} \rangle = \frac{1}{2}\epsilon \left[ \frac{\alpha-\gamma}{1+\omega\tau^2}(\nabla T \cdot \boldsymbol{E}_{ac})\boldsymbol{E}_{ac} - \frac{1}{2}\alpha|\boldsymbol{E}_{ac}|^2 \nabla T \right] (3),$$

here $\epsilon$ is the fluid permittivity and $\sigma$ is the fluid conductivity; $\alpha = (1/\epsilon)(\partial\epsilon/\partial T)$ and $\gamma = (1/\sigma)(\partial\sigma/\partial T)$; $\tau = \epsilon/\sigma$ is the charge relaxation time; $\boldsymbol{E}_{ac}$ is the local AC electric field.

In an early work by Ndukaife *et al.* (Figure 4(a)) [26], they demonstrated that the ETP flow can be generated from a plasmonic resonator made of a gold nanopillar. The gold nanopillar also served as plasmonic nanotweezers. The ETP flow transported a nanoparticle to the vicinity of the nanopillar to rapidly load the plasmonic nanotweezers within seconds. Their results highlighted the rapid transport velocity and long-distance operation range of the ETP flow. These advantages of ETP flows also benefit biosensor platforms to actuate detection at lower analyte concentrations with a shorter incubation timespan [103,104]. Many works have been reported establishing ETP flows to transport and manipulate nanosized entities [26,100,101,104–108] and some of them are summarized in Figures 4(a) to (f). Gold pillar arrays, patterned metal films, double nanohole apertures, and even dielectric photonic crystals have been utilized to achieve ETP flow for rapid particle transport. For the case of dielectric photonic crystal cavities, the ETP flow comes from the local heating of the water layer near 1550 nm wavelength where water absorbs [105].

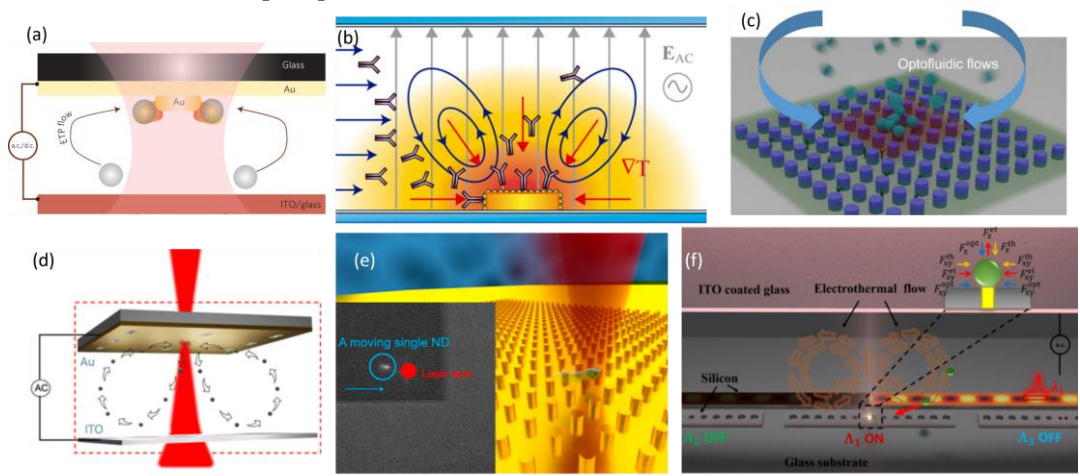

**Figure 4**: Examples of eletrothermoplasmonic flow and its applications in biosensing. (a) demonstrates the generation of ETP flows from a gold nanopillar array to rapidly load the near-field plasmonic nanotweezer of the nanopillar. (a) adapted with permission from ref [26]. Copyright 2016 Springer Nature. (b) shows that a biosensing platform is integrated with ETP flow to speed up the incubation process. The biosensor is made of an array of gold nanodisks. (b) adapted with permission from ref [103]. Copyright 2018 American Chemical Society. (c) also utilizes ETP flows to expedite the incubation process of an image-based extracellular vesicle sensing platform. (c) adapted with permission from ref [104]. Copyright 2023 ARXIV. (d) uses double nanohole aperture instead of embossed nanopillars to perform ETP-assisted trapping. (d) adapted with permission from ref [106]. Copyright 2023 Royal Society of Chemistry. (e) demonstrated that at low AC frequency, ETP flow can work together with other electrohydrodynamic effects to capture and manipulate a single quantum emitter on a nanopillar array. (e) adapted with permission from ref [108]. Copyright 2021 American Chemical Society. (f) shows a theoretical work discussing the generation of electrothermal flows in an all-dielectric photonic crystal platform. The temperature rise is mainly from the water absorption arising from the extremely localized electric field in a photonic crystal cavity. (f) adapted with permission from ref [105]. Copyright 2023 American Physical Society.

### 2.2.2 Opto-thermo-electrohydrodynamic tweezers

One requirement of the electrothermoplasmonic tweezers mentioned in the last subsection is that it leverages plasmonic heating and the trapped nanoparticles are always near the thermal hotspot where the local temperature rise is high. While this does not present impediments when handling synthetic colloidal particles, it can be undesirable for certain applications that require the manipulation of delicate biological molecules.

One strategy to overcome this limitation demonstrated by Hong *et al.* [109] showcases a 'stand-off' trapping platform, called opto-thermo-electrohydrodynamic tweezers (OTET), as shown in Figures 5(a) and (b). On this platform, a gold film patterned with a nanohole array is used as the electrode with an AC voltage applied perpendicular to the pattered electrode across a microfluidic chamber. This patterned gold film generates **AC electroosmotic (ACEO)** flows [98]. The generation of ACEO flow can be explained as following. Near the interface of the electrode and liquid, the ions within the liquid form a double-layer structure, known as the electric double layer (EDL). EDL screens the surface charge of the electrode. When the electrode itself is patterned with nanostructures, the local AC electric field is perturbed and results in tangential components parallel to the electrode surface. These tangential electric field components drive ions within the EDL to move along the electrode surface, initiating fluid motion known as ACEO flow [102,110].

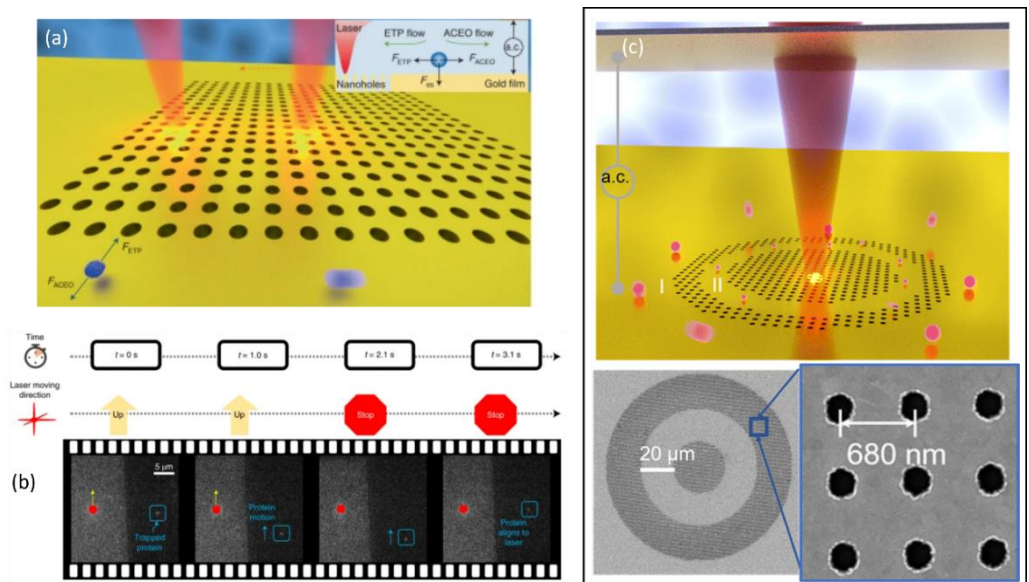

**Figure 5**. Examples of opto-thermo-electrohydrodynamic tweezers (OTET) leveraging the interplay between ETP flows and low-frequency ACEO flows. OTET gently traps a single biological molecule, or a nanoparticle, away from the laser illumination (also the high-temperature region), but still enables dynamic manipulation of the trapped particles "at a distance". (a) and (b) demonstrate the "stand-off" trapping and manipulation of a single bovine serum albumin protein molecule by OTET. (a) and (b) adapted with permission from ref [109]. Copyright 2020 Springer Nature. (c) shows a design stemming from OTET that enables size-based trapping of extracellular vesicles. The trapping stability is enhanced when the particles are trapped within the region II. (c) adapted with permission from ref [111]. Copyright 2023 Royal Society of Chemistry.

In the OTET system, a laser was illuminated on the nanohole array to generate photo-induced heating efficiently. As the AC electric field is applied, ETP flow is also induced. In the radial direction, the ETP flow is directed toward the laser illumination near the patterned gold film, while the ACEO flow is pointing radially outward in the opposite direction of the ETP flow. Under a low AC frequency (below 10 kHz for instance), the strength of ACEO flow balances with the strength of ETP flow to create stagnation zones outside the nanohole array, where the nanoparticles would be trapped at the stagnation zones. Ultimately, the interplay between ACEO flows and ETP flows ensures that the nanoparticles (sub-10 nm protein molecules, for example) are trapped away from the thermal hotspot to preclude photo-induced heating. OTET also allows for dynamic manipulation by translating the laser spot on the nanohole array and size-based sorting by tuning the AC frequency. OTET provides a new paradigm for single nanosized particle manipulation and analysis in a safe and efficient manner.

By tuning the geometry of the nanohole array, Figure 5(c) shows that another type of high stability opto-thermo-electrohydrodynamic tweezers for stably trapping, manipulating and size-based sorting of single extracellular vesicles has also been achieved [111].

### 2.2.3 Optically-induced dielectrophoretic manipulation

Electrokinetic effects are also advantageous in manipulating colloidal particles in parallel simultaneously, improving the efficiency of manipulations. For example, **dielectrophoresis** (DEP) is one of the widely adapted approaches for electrokinetically manipulating microscopic objects. DEP forces are similar to optical forces, whose magnitude is proportional to the gradient of the electric field intensity, and the volume of the trapped particles [112]. Therefore, positive DEP force allows particles to be trapped at the position where the electric field line has a high density, such as a sharp edge of an electrode [113–116]. On the other hand, negative DEP expels particles from the high electric field region.

To enable dynamic manipulation, optoelectronics tweezers (OET) [22,117–120] have been demonstrated to harness optically-induced dielectrophoresis to manipulate microparticles and cells. The working principle of OET can be explained by the scheme shown in Figure 6(a). When the a-Si:H layer is exposed under light illumination, the electrical conductivity of the film changes due to photoconductivity, thereby creating 'virtual conductors' at the illuminated regions on-chip. The a-Si:H film sits on an ITO electrode, and the nonuniformity of electrical conductivity alters the local electric field, establishing a field gradient to induce DEP. The location of the DEP force can be controlled on demand by the projected light pattern. Here, in Figure 6(a), the authors utilized negative DEP to trap the polystyrene bead inside the ring-shape pattern, where the shape of the pattern can be arbitrarily controlled by a Digital Micromirror Device (DMD) [117]. Using a well-tailored light pattern matching the periphery of a micro-gear, Zhang *et al.* demonstrated a micro-spanner driven by OET [120]. The schematic illustration is depicted in Figure 6(b).

Several works have also shown engineering the photoconductive substrate to improve the performance of OET. Zhang *et al.* demonstrated the patterned a-Si:H film can still generate optically-induced DEP [119]. The patterned-OET platform, as shown in Figure 6(c), provided an extra functionality to repel unwanted microparticles/cell while keeping only the target particle. Apart from a-Si:H, other photoconductive materials, such as titanyl phthalocyanine pigments (TiOPc) [121,122] or bulk heterojunction polymers [123,124], were also explored to be adapted in OET. Figure 6(d) exemplifies the use of (TiOPc) to induce DEP and to manipulate a microbubble in the channel [122]. The photoconductive polymers are easy to fabricate and provide strong light absorption over a wide spectral range.

As above-mentioned, unfortunately, OET faces challenges for nanoscale particle manipulations, because the magnitude of DEP force drops significantly as the trapped particle size reduces. Although it is difficult to directly use DEP force to stably trap nanoscale particles, one way to harness DEP in nanomanipulation is to combine it with plasmonic nanotweezers. The conductive nature of plasmonic materials makes them intrinsically good candidates as the electrodes. The main trapping force is contributed by the optical gradient force from the plasmonic nanocavity. Babaei *et al.* demonstrated DEP-assisted double-nanohole tweezers to expedite the occurrence of the trapping events [116], where DEP was reported to enable bringing nanoparticles close to the double-nanohole cavity.

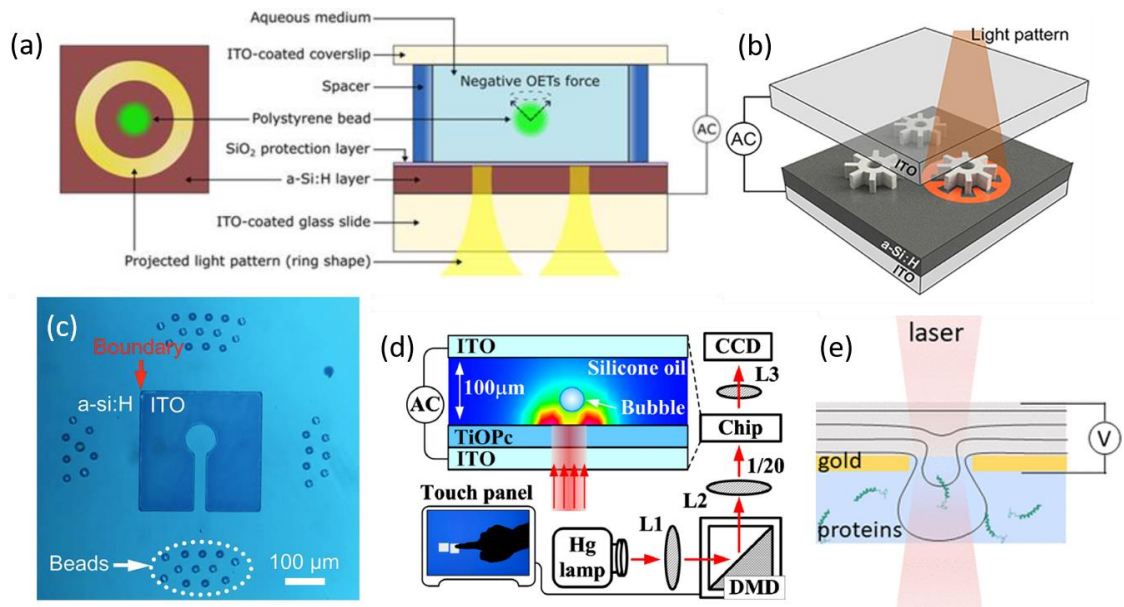

**Figure 6**: Examples of optically-induced dielectrophoretic manipulations. (a) shows a design of optoelectronic tweezers utilizing the photoconductivity of amorphous silicon. (a) adapted with permission from ref [117]. Copyright 2023 AIP Publishing. (b) presents a micro-spanner based on OET to rotate a microsized gear. (b) adapted with permission from ref [120]. Copyright 2021 Springer Nature. (c) demonstrates an OET platform comprised of a patterned a-Si:H film, where it inherently repels unwanted microbeads. (c) adapted with permission from ref [119]. Copyright 2018 Wiley-VCH. (d) manifests an OET system made of photoconductive polymer, TiOPc. (d) adapted with permission from [122]. Copyright 2011 AIP publishing. (e) demonstrates that fringe-DEP can expedite the transfer of protein molecules to double nanohole nanotweezers. (e) adapted with permission from ref [125]. Copyright 2023 American Chemical Society.

### 2.2.4 Geometry-induced Electrohydrodynamic Tweezers benefit plasmonic nanotweezers

As noted previously, a long-standing challenge with plasmonic nanotweezers concerns how to deliver suspended particles to the plasmonic cavities without relying on slow Brownian diffusion. Hong *et al.* recently proposed and demonstrated a novel concept of the geometry-induced electrohydrodynamic tweezers (GET) [98], which is a perfect manifestation of how the long-range parallel-manipulating nature of electrohydrodynamic effects assists the near-field plasmonic nanotweezers.

The GET trap demonstrated rapid and massively parallel trapping of single extracellular vesicles using ACEO flows. The electrohydrodynamic effects produced by the patterned gold film accelerated the trapping process and enabled self-limited single extracellular vesicle trapping. Figure 7(a) schematically illustrates that this GET platform consists of a gold film patterned with radial arrays of nanoholes. The EVs are rapidly transported and electrohydrodynamically trapped by the ACEO flows generated by the nanohole array. At the center of each trapping region, a double nanohole aperture was milled, serving as plasmonic nanotweezers when illuminated by a laser. Figure 7(b) depicts the simultaneous parallel trapping of multiple extracellular vesicles on the GET traps using ACEO flows in an AC electric field.

The step-by-step manipulation process, depicted in Figure 7(c), includes: (1) capturing multiple single EVs; (2) employing a laser to selectively hold an EV within a plasmonic cavity; (3) deactivating the AC electric field to release unselected EVs; and (4) ultimately freeing the optically trapped EV. This system marks a pioneering achievement in swiftly loading EVs into a plasmonic cavity without generating detrimental heating effect, a feature critical for its application in delicate biological analyses, such as studying individual extracellular particles to understand their heterogeneity. The entire setup is mounted on a sapphire substrate to dissipate excess heat, ensuring a negligible temperature rise.

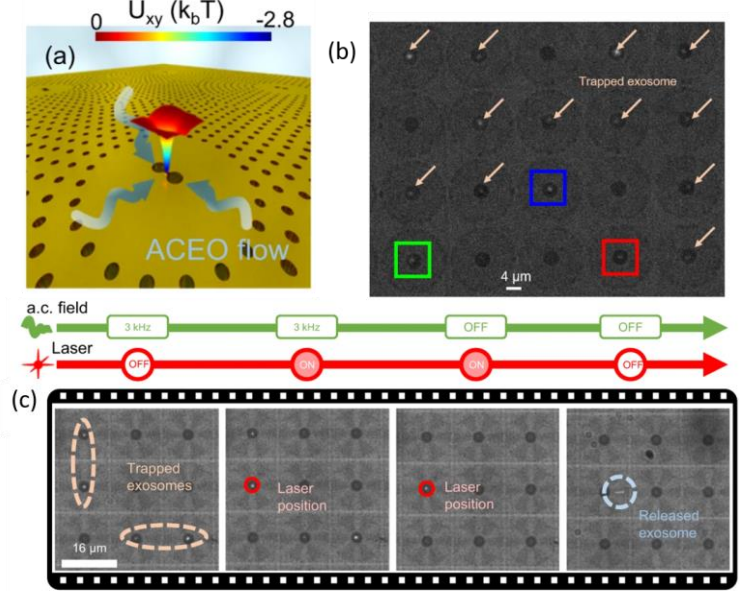

**Figure 7**: Geometry-induced electrohydrodynamic tweezers (GET). (a) schematically depicts working principle of GET, where ACEO flows transport the nanoparticles towards the center of the trap, near the plasmonic cavity (double nanohole). (b) demonstrates the trapped EVs in parallel by GET. (c) demonstrates the complete operation process of GET. (a) to (c) adapted with permission from ref [98]. Copyright 2023 Springer Nature.

Up to now, we have reviewed the combined effects of light and electrohydrodynamics. They enable us to load the traps rapidly, to achieve dynamic manipulation, and to capture nanoparticles far away from traps in low-concentration solutions. In the following sections, we will review other emerging nanomanipulation methods that rely more on light-induced heating. Many of them leverage the properties of colloidal particles, the surface charge on the microfluidic chamber wall, or the modified nanoparticle properties when adding surfactant.

### 2.2.5 Opto-thermophoretic tweezers

Thermophoresis describes the movement of colloidal entities within a fluid under a thermal gradient [126,127]. As a colloidal particle exists in a thermal gradient field, the fluid surrounding the particle experiences nonuniform temperature distribution and unbalanced hydrostatic pressure emerges on each side of the particle. This pressure difference drives the particle moving towards the direction with reducing interfacial free energy [128,129]. In general, the migration velocity from thermophoresis is governed by:
$$u_{thermophoresis} = -D_T \nabla T \quad (4)$$
Here $D_T$ is the thermophoretic mobility. $D_T$ can be calculated using the Soret coefficient $S_T$:
$$D_T = S_T D \quad (5)$$
Here $D$ is the Brownian diffusion coefficient. We can easily conclude from the equation that $S_T$ decides the moving direction of the particle. When $S_T > 0$ (*i.e.,* positive thermophoresis), the particle moves from the hot region to the cold region. When $S_T < 0$ (*i.e.,* negative thermophoresis), the particle migrates from the cold region to the hot region. In general, most colloidal particles have positive $S_T$, meaning they are prone to move away from the heat source. This is not favorable in a plasmonic near-field tweezers since it suggests that the photo-induced heating prevents particles from getting close to the plasmonic nanostructures to be trapped [90].

However, there are also strategies utilizing positive thermophoresis for trapping as shown in Figures 8(a) and (b). Figure 8(a) shows a configuration where plasmonic nanoantennas are assembled in a circular array. The laser was swept over the plasmonic chain to introduce continuous positive thermophoresis and push nanoparticles towards the center for trapping [130,131]. Another strategy shown in Figure 8(b) utilized laser-cooling technique, where the low-loss upconverting material reduces the local temperature upon laser illumination [132].

On the contrary, Figure 8(c) describes that certain biological cells (yeast cells and Escherichia coli cells) exhibit a negative Soret coefficient, making them suitable for manipulation with negative thermophoresis [133]. Lipid vesicles are also found to be able to overcome positive thermophoresis with low concentrations of NaCl added into the solution, which is presented in Figure 8(d) [134]. When the environmental temperature is cooled down from room temperature, the Soret factor can be tuned to a negative value to initiate negative-thermophoretic trapping. Figure 8(e) shows a trapping platform based on this phenomenon, called 'hypothermal opto-thermophoretic tweezers [135]. Polystyrene beads are also demonstrated to be trapped and manipulated under interfacial-entropy-driven thermophoretic force when adding surfactants into the solutions to modify the surface charges on polystyrene beads [128]. Similarly, Figure 8(f) experimentally investigated the impact of thermophoresis on the trapping stability of plasmonic nanotweezers when various surfactants were added into the solutions [92]. It is worth noting here that a unifying physical model of thermophoresis has yet to be established [126,127,136–139].

Thermophoresis has also been combined synergically with **Buoyancy-driven convection**. They have been shown to facilitate biological particle aggregation [140–143]. Convection originates from the density gradient of the fluid in a temperature gradient field. Temperature rise normally reduces the density of fluid, making the hot fluid move upward vertically due to the Archimedes' force [144,145].

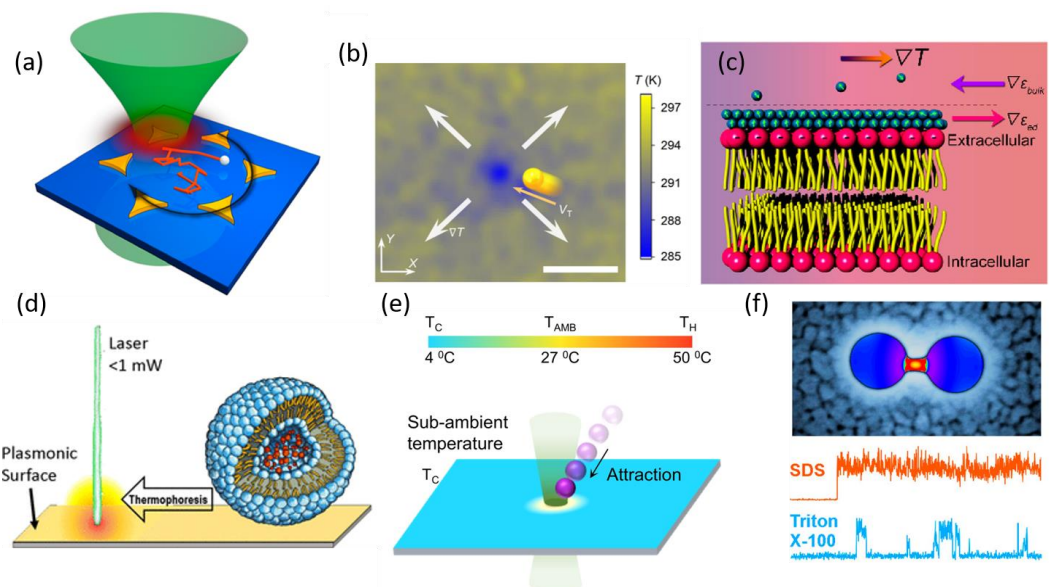

**Figure 8**. Optically-induced thermophoretic manipulations. (a) shows a trapping platform leveraging positive thermophoresis. A laser illuminates the array of gold nanotriangles and endows positive thermophoresis to push nanoparticles inside the array. (a) adapted with permission from ref [131]. Copyright 2013 American Chemical Society. (b) demonstrated a photo-induced cooling technique leveraging anti-stokes shift of the upconverting material, so that positive thermophoresis pushes the particle towards the laser illumination region. (b) adapted with permission from ref [132]. Copyright 2021 AAAS. (c) shows the trapping and manipulation of certain types of cells that were found to exhibit negative thermophoresis in water. (c) adapted with permission from ref [133]. Copyright 2017 American Chemical Society. (d) demonstrated that a lipid vesicle can be manipulated by opto-thermophoretic attraction. A small amount of NaCl was added to the solution to overcome the positive thermophoresis. (d) adapted with permission from ref [134]. Copyright 2018 American Chemical Society. (e) showcases another way to introduce negative thermophoresis, which is to cool down the environmental temperature below room temperature. (e) adapted with permission from ref [135]. Copyright 2023 Springer Nature. (f) presents a quantitative analysis of the influence of surfactant and thermophoretic force on plasmonic nanotweezers. (f) adapted with permission from ref [92]. Copyright 2020 American Chemical Society.

### 2.2.6 Opto-thermoelectric-mediated manipulation

When certain types of surfactants are added into a particle solution, such as cetyltrimethylammonium chloride (CTAC), thermoelectric effect can take place [146,147]. As the CTAC concentration surpasses the critical micelle concentration, $CTA^+$ groups start to aggregate and form CTA micelles that are positively charged. Subsequently, the $CTA^+$ groups coat the negatively-charged surface of the suspended colloidal particles due to the Coulombic interaction, resulting in overall positively-charged particles, in addition to the negatively charged chlorine ions in the solution. In the temperature gradient, micelles, chlorine ions, and colloidal particles start to migrate away from the heat source due to positive thermophoresis. It turns out that the migration velocities of different ions/particles are different. Micelles migrate much faster than Chlorine ions. Consequently, at steady state, the micelles are statistically distributed further away from the heat source than chlorine ions. This spatial separation of positively and negatively charged ions results in the establishment of an electric field, called thermoelectric field, also known as the Seebeck effect [137,148]. Such thermoelectric field points towards the heat source and its magnitude can be expressed as

$$E_{thermoelectric} = \frac{k_B T \nabla T}{e} \frac{\sum_i Z_i n_i S_{Ti}}{\sum_i Z_i^2 n_i} \quad (6)$$

Here $T$ is the environmental temperature, $e$ is the elementary charge, $i$ stands for the ionic species. For each species, $Z$ represents the charge number, $n$ represents the concentration of the species and $S_T$ represents the corresponding Soret coefficient. The positively charged colloidal particles migrate towards the heat source following the electric field line. A thorough study on the quantitative analysis of thermoelectric effects has been published by Kollipara *et al.* [147].

In experiments, the heat for thermoelectric field can be easily generated by illuminating a thin metal film with a focused laser beam, which has been shown by Lin et al. to demonstrate thermoelectric field-induced manipulation of single metal nanoparticles with various shapes, including nanospheres, nanorods, or nanotriangles made of silver or gold [146]. The so-called 'opto-thermoelectric tweezers' are also not limited to single nanoparticle trapping. The authors demonstrated the assembly of multiple nanoparticles by reshaping the laser profile. They also employed dark-field spectroscopy to characterize the size of the trapped nanoparticles and to monitor the plasmonic coupling between nanoparticles when multiple particles were trapped together. A follow-up paper from the same group [149] presented thermoelectric field-induced trapping of nanoparticles comprising of different materials, including metals, polystyrenes, or quantum dots. Figure 9(a) shows the schematic of another work, where a nanoradiator made of plasmonic bars was used to replace the gold film as the heat generator. By placing the nanoradiators into an array, particle transport between nanoradiators can be controlled by shifting the polarization states of laser illumination.

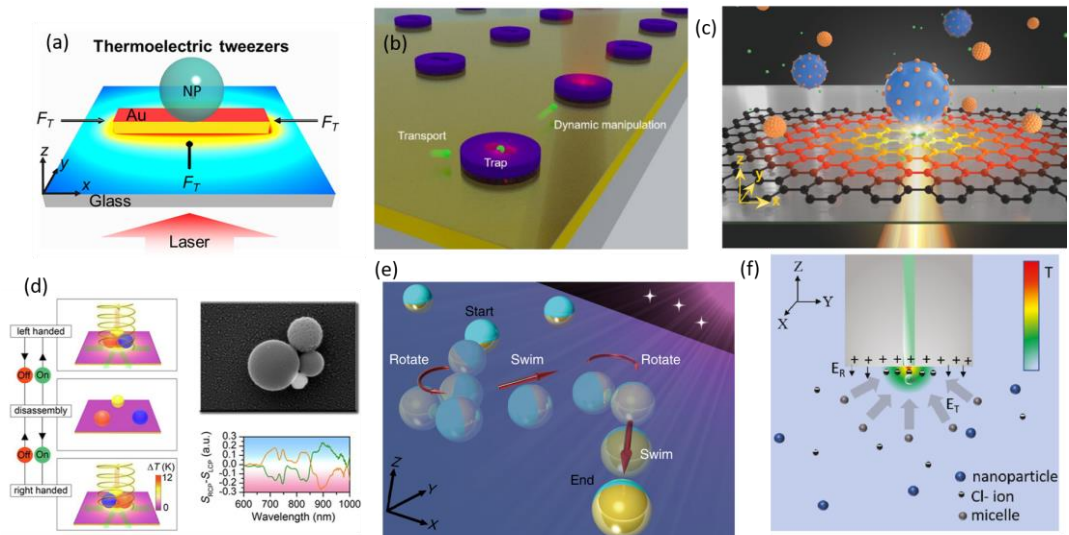

Figure 9: The thermoelectric effect has been widely utilized in manipulating particles. (a) demonstrates the idea of using a plasmonic nanoradiator to generate heating efficiently. The induced thermoelectric effect was then used to trap

nanoparticles. (a) adapted with permission from ref. [149]. Copyright 2018 American Chemistry Society. (b) presents the optical-anapole-mediated thermoelectric trapping and manipulation of single extracellular vesicles. (b) adapted with permission from ref. [150]. Copyright 2023 ARXIV. (c) demonstrated opto-thermoelectric tweezers based on graphene. (c) adapted with permission from ref. [151]. Copyright 2021 Wiley-VCH. (d) proposed and demonstrated the assembly and manipulation on a chiral 'meta-molecule' in a reconfigurable manner using opto-thermoelectric effect. (e) adapted with permission from ref. [152]. Copyright 2019 Elsevier. (e) shows Janus nanoparticles can be manipulated with various motion modes enabled by the opto-thermoelectric effects due to the asymmetric absorption from each side of the particle. (e) adapted with permission from ref. [153]. Copyright 2020 Springer Nature. (f) presents the integration of opto-thermoelectric tweezers on the tip of a fiber. (f) adapted with permission from ref. [154]. Copyright 2019 De Gruyter.

Figure 9(b) demonstrates a hybrid dielectric-plasmonic trapping system enabled by optical anapoles [150]. The optical anapole provides a strong electromagnetic near-field enhancement for optically trapping a single extracellular vesicle. At the same time, the plasmonic reflector underneath it further boosts the field enhancement and provides photo-induced heating to initiate the opto-thermoelectric effect. This work demonstrated rapid particle transport and dynamic manipulation enabled by the opto-thermoelectric effect. Figure 9(c) presents graphene-based opto-thermoelectric tweezers, [151], that employed graphene instead of relying on traditional plasmonic materials. The authors also demonstrated the ability to optically pattern the graphene layers.

Not limited to the nanoparticles with regular geometries, the opto-thermoelectric effect has been shown to assemble nanoparticles with different sizes and materials (metal, polystyrene, or Si) into a customized 'meta-molecule' [152], which is illustrated in Figure 9(d). Due to this size and material inhomogeneity, a circular dichroism signal was detected from the light scattered by the meta-molecules, indicating the chiral property of this artificial 'molecule'. This chiral meta-molecule can also be reconfigured by disassembly and reassembly.

In addition to engineering the substrate to generate photo-induced heating, absorption from the nanoparticles has also been utilized to actuate thermoelectric effect [153,155]. For instance, Figure 9(e) shows the scheme of the 'opto-thermoelectric microswimmer' made of Janus particles. A Janus particle is a half-dielectric, half-metallic particle. Thus, its absorption of light is different on each side. Leveraging this temperature gradient on the surface of Janus particles, Peng *et al.* [153] demonstrated the optothermophoretic microswimmer, where the Janus microparticle has two states of motion: 'swim' and 'rotate'. The thermoelectric force from the temperature gradient on the microparticle surface drives the microparticle moving forward, namely the 'swim' state. The thermal fluctuation on the Janus particle randomly changes the orientation of the particle, and results in the 'rotation' of the Janus particle. A precise steering of the microparticle moving direction can be controlled by a feedback system.

Other applications of opto-thermoelectric effects, including nano-pipette [154] or particle filtration [156], have also been achieved. Figure 9(f) shows an example of integrating the opto-thermoelectric tweezers onto a fiber tip for delivering nanoparticles into lipid vesicles [154].

### 2.2.7 Optically-induced phoretic manipulations

Adding solute into the liquid can also bring about **diffusiophoresis**, where the solute can be either ionic or nonionic. In ionic solutions, the diffusiophoretic forces stem from the biased charge distribution in the EDL, causing electric and pressure field gradients. In a nonionic solution, diffusiophoresis is solely driven by the pressure field gradient. Diffusiophoresis is generally influenced by three factors [126]: (1) the concentration gradient of the solute; (2) the surface charge of the colloidal particles; and (3) the diffusion coefficient of electrolytes.

Depletion attraction is a special kind of diffusiophoresis. In optothermal nanomanipulation, the photo-induced heating causes positive thermophoresis of solutes so that the concentration of the solutes near the hot spot is smaller than the concentration further away from the hot spot. Such induced solute concentration gradient then exerts the diffusiophoretic (depletion attraction) forces onto target colloidal particles. The target colloidal particles, therefore, tend to move towards the hotspot.

Several demonstrations have been conducted on the utilization of opto-thermal diffusiophoresis to manipulate diverse types of nanoparticles, such as Janus particles (Figure 10(a)), living cells (Figure 10(b)), DNAs (Figure 10(c)), or polystyrene beads (Figure 10(d) and (e)). Figure 10(a) shows the scheme of an opto-thermal diffusiophoretic microswimmer based on a Janus particle [157]. Lutidine was added into the Janus particle solution, where the concentration of lutidine was well-tailored to determine the critical temperature $T_c$. As the Janus particle was illuminated with light, the absorber-coated side generated heat to 'demix' lutidine on one side of the Janus particle. It thus leads to a concentration gradient of lutidine locally, and the Janus particle was driven by self-diffusiophoresis.

The choices of solutes also cover a wide range including polyethylene glycol (PEG), NaCl, KCl, and other types of ionic or nonionic solutes [96,158–164]. Among them, PEG is widely chosen because of its superb bio-compatibility. Figure 10(b) shows the example of using opto-thermal diffusiophoresis to manipulate and assemble DNAs or living cells in PEG solutions [163]. The interplay between the thermophoresis of DNAs and PEG-mediated depletion force was also studied in the work highlighted in Figure 10(c).

The heat source is also not limited to simple metal films. Multiple types of nanoantennas are also demonstrated to generate heat efficiently for inducing opto-thermal diffusiophoresis, such as plasmonic nanoantennas (Figure 10(d)) [162] or dielectric nanostructures. These nanoantennas can be carefully designed to provide additional optical gradient force to enhance trapping. For example, Hong et al. (Figure 10(e)) [164]. have experimentally demonstrated the trapping, transportation, and manipulation of nanoscale particles in PEG-containing media. In this setup, a 532 nm continuous laser was loosely focused on the nanoantenna, which is designed to excite the anapole state. The nanoantenna generated local heating due to the inherent loss from silicon under the excited wavelength of light. The locally increased temperature caused the PEG molecules to move to the colder regions due to the positive thermophoretic behavior of PEG molecules. The repelling action toward the cold region of PEG established a gradient of PEG concentration [165]. The nanoparticles then began to move towards the hot region where the anapole nanoantenna was located, as the gradient of PEG concentration induced the diffusiophoretic force (depletion attraction force). Once the nanoparticle reached the vicinity of the anapole whereby the near-field started to have an effect, the optical gradient force contributed additionally to trapping the nanoparticles. In addition, the depletion attraction between colloidal nanoparticles and the substrate also allows nanoparticles to 'be printed' onto the substrate [166,167].

Photophoresis is another kind of optically-induced phoretic force which describes the directed movement of light-absorbing objects suspended in either aqueous or gaseous conditions [126]. It is attributed to the momentum transfer between the object and the surrounding media molecules. As an application, optical pulling of particles in gases has been reported. The absorbing objects in a light field are prone to move from the high-intensity region to the low-intensity region, thus structured light has been proposed and designed to leverage photophoretic force to trap absorbing particles in the low-intensity region, such as at the center of a donut beam [168]. In addition, Lu *et al.* investigated the manipulation of a micron-size gold plate on the tip of a tapered fiber by managing the interplay between the optical force and the photophoretic force [169], as illustrated in Figure 10(f). The gold microplate can be pulled or pushed back and forth on the tapered fiber with supercontinuum light in ambient air.

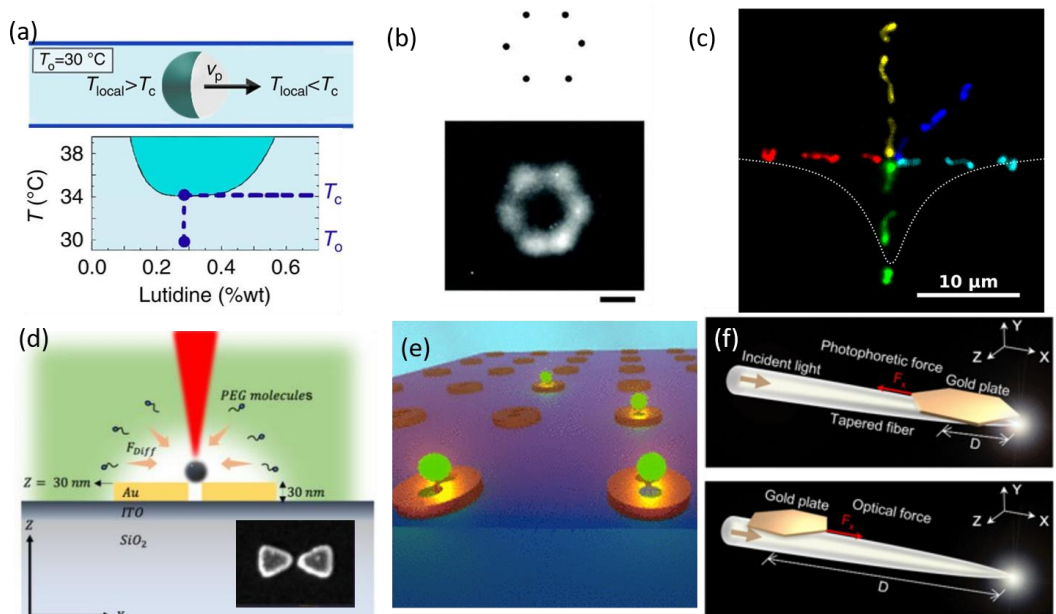

**Figure 10:** Examples of optical manipulation based on diffusiophoresis or photophoresis. (a) presents the scheme of a self-diffusiophoretic microswimmer made of a Janus particle. (a) adapted with permission from ref [157]. Copyright 2016 Springer Nature. (b) shows the versatile manipulation of soft biological materials by opto-thermal diffusiophoresis. (b) adapted with permission from ref [163]. Copyright 2013 AIP Publishing. (c) illustrates the work studying the interplay between the thermophoresis of DNAs and PEG-induced depletion interaction. (c) adapted with permission from ref [170]. Copyright 2023 American Chemical Society. (d) and (e) both leverage the photo-induced heating from nanoantennas to actuate the opto-thermal diffusiophoretic motion of nanoparticles. PEG was added into the solutions to provide the concentration gradient. The difference between these two works is that (d) used a plasmonic bowtie resonator, while (e) employed the loss of amorphous silicon at visible wavelengths, providing a new perspective of studying thermo-nanophotonic in dielectric materials. (d) adapted with permission from ref [162]. Copyright 2023 ARXIV. (e) adapted with permission from ref [164]. Copyright 2023 American Chemical Society. (f) shows a pulling or pushing manipulation mechanism enabled by the synergic effect of optical force and photophoretic force. (f) adapted with permission from ref [169]. Copyright 2017 American Physical Society.

### 2.2.8 Optically-induced manipulation with thermo-osmotic flows

Thermo-osmosis is imposed by a nonuniform heat distribution along the solid-liquid interface [171-173]. Along the direction tangential to the surface, the temperature gradient induces excessive free energy and, thus, excessive enthalpy in the very vicinity of the solid surface. This enthalpy difference finally causes a liquid flow from the cold region to the hot region, termed thermo-osmosis. On the surface of a Janus particle, the generated temperature difference of each side of the Janus particle under light illumination induces osmotic pressure on the particle surface, which has been utilized for designing microswimmers or rotors [174–176]. Figure 11(a) and (b) present two examples of manipulations driven by this self-osmotic pressure.

Many other works have also studied the thermo-osmotic flows near a flat solid substrate [172,173,177,178]. Thermo-osmosis exerts confinement onto particles in the tangential direction and is especially useful for manipulating metal nanoparticles, as shown in Figure 11(c) [172]. This is because metal particles have higher thermal conductivities and are less impacted by thermophoresis.

Yang *et al.* introduced a synergistic strategy for controlling nanoparticle clusters by leveraging the optofluidic interplays between thermo-osmotic flows, thermophoresis, and convection flows near an all-dielectric metasurface [143]. As shown in Figure 11(d), this design employs a quasi-BIC-driven all-dielectric metasurface to exert subwavelength-scale control over temperature and control fluid motion. In this system, the optofluidic behaviors are influenced by the interaction between the substantial water absorption and the intense absorption from the water layer adjacent to the quasi-BIC metasurface. The pronounced electromagnetic field enhancement at the quasi-BIC resonance significantly increases the flow velocity—up to three times higher than in the off-resonant conditions—by simply adjusting the wavelength within a range of several nanometers.

Another application that synergistically utilizes thermo-osmosis and depletion attraction is presented in Figure 11(e). This work demonstrated the enrichment of DNA-conjugated 80 nm gold nanospheres due to the negligible thermophoresis experienced by the gold nanospheres. The CRISPR-related proteins and DNA strands were also concentrated near the hotspot. This platform thus sets up an enhanced CRISPR-based single-nucleotide polymorphism detection at the single molecule level [179].

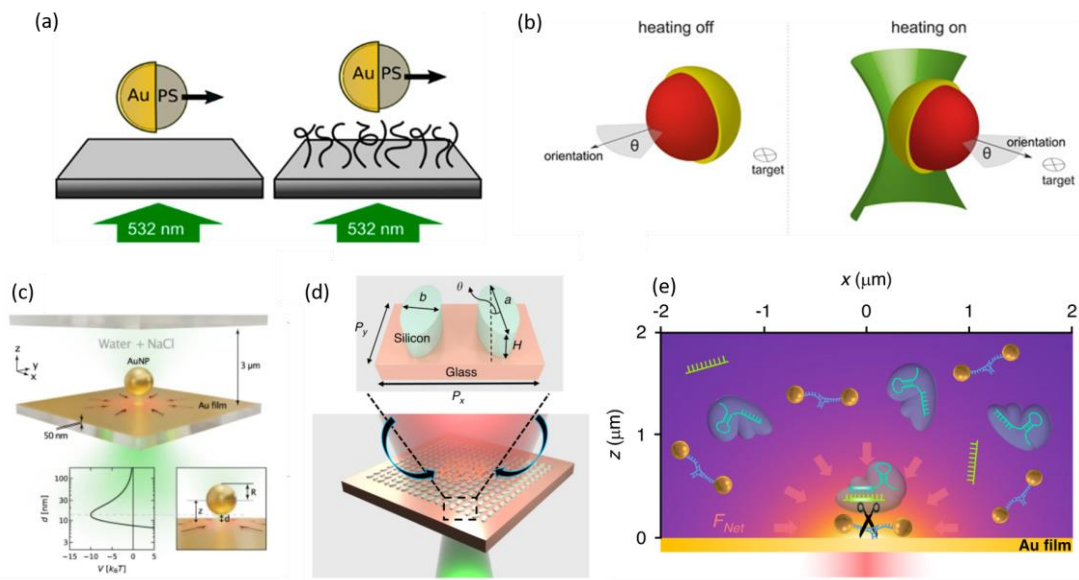

**Figure 11**: (a) shows a study of the self-propulsion of Janus particles near a functionalized substrate. It investigates the impact of the substrate-particle interaction on particle motion when self-osmotic pressure occurs on the Janus particle. (a) adapted with permission from ref. [175]. Copyright 2020 American Chemistry Society. (b) presents the adaptive steering and localization process of a self-propelled microswimmer by 'photon nudging'. (b) adapted with permission from ref. [174]. Copyright 2023 Springer Nature. (c) presents the trapping of gold nanospheres using thermo-osmotic flow. (c) adapted with permission from ref. [172]. Copyright 2022 Springer Nature. (d) shows the scheme of optofluidic transport and assembly of nanoparticles using an all-dielectric quasi-BIC metasurface. (d) adapted with permission from ref. [143]. Copyright 2023 Springer Nature. (e) demonstrates an application of using the combination of depletion attraction and thermo-osmosis for conducting CRISPR. (e) adapted with permission from ref. [179]. Copyright 2023 Springer Nature.

These hybrid effects, including other important methods but not elaborated in this chapter, such as Marangoni convections [180–182] or optothermal deformations [183,184], overcome the intrinsic limitations of optical tweezers and provide new perspectives on the advancement of light-based manipulations.

## 3. Optofluidic sensors

For the traditional chemical or biological sensors, a key challenge lies in the slow diffusion of analytes to the sensing area, preventing rapid real-time detection and analysis. To achieve rapid and high sensitivity sensing, researchers have merged optics and microfluidics to create versatile lab-on-a-chip optofluidic platforms for analyzing samples within a microfluidic environment. Optofluidic devices are notable for their portability, repeatability, sensitivity, speed, high-throughput, and cost-effectiveness. For example, optofluidic devices can leverage photothermal effects to induce controlled thermal gradients that enable ETP flow-assisted transport of analytes to the sensor region within seconds. Furthermore, the multiple fluidic channels in optofluidic devices support multiplexed measurements, enabling simultaneous analysis of various samples. Optofluidic sensors have been at the forefront of research since the field's inception in 2003 [6,185–188]. Various optical technologies, such as refractometry, colorimetry, surface enhanced Infrared spectroscopy (SEIRA), surface-enhanced Raman scattering (SERS), fluorescence, phase interrogation, etc have been employed, either individually or in tandem, to enhance the sensing process [186–188]. In this section, we discuss the state-of-art optofluidic sensors utilizing various optical properties and elaborate on how microfluidics technology helps improve the performance of the sensors.

## 3.1 Optofluidics-enabled label-free refractive index sensor
### 3.1.1 Plasmonics-assisted optofluidic sensors

A significant subset of sensors achieve label-free sensing by employing plasmonic structures, as the surface plasmon resonance (SPR) and localized surface plasmon resonance (LSPR) in plasmonic structures are extremely sensitive to refractive index changes of the surrounding environment. However, the transport of analytes to the plasmonic hotspots, real-time monitoring, surface cleaning and reusability are difficult for refractometry setup [186]. To address these challenges, researchers have combined microfluidic technology and plasmonic nanostructures, developing various advanced optofluidic plasmonic sensors. For instance, shown in Figure 12(a), Fan et al. [189] integrated nine microfluidic channels with gold nanoparticles on a smartphone platform. The system, featuring 81 sensor units distributed among the channels, is capable of identifying various biomarkers in a 100 µL of sample volume through the camera. Instead of using nanoparticles, Coskun et al. [190] fabricated gold nanohole arrays for biosensing application. Shown in Figure 12(b), the microfluidic channel is utilized for sample introduction, precise delivery of analytes and sensing. Similarly, Li et al. [191] utilized gold nanohole arrays with microfluidic cell culture systems, achieving real-time, label free monitoring of live cell secretions. This approach overcomes the limitations of conventional methods like ELISA, offering more dynamic, continuous, and direct analysis of cellular activities. Im et al. [192] presented a plasmonic hole array sensor integrated with 12 microfluidic channels to detect and profile exosomes for cancer diagnostics. The microfluidic channels not only enable controlling small volumes of exosome samples but also facilitate the functionalization of the sensor surface and surface cleaning. In Figure 12(c), Chen et al. [193] developed a microfluidic sensor device for multiplex immunoassays of six cytokines in a complex serum matrix. They integrated 480 nanoplasmonic sensing spots within the microfluidic channel arrays, enabling detecting concentration as low as 20 pg/µL from 1 µL serum sample. Hong et al. [104] recently achieved the rapid transport and enrichment of EVs through ETP flows generated by a dielectric metasurface on a thin layer of gold coating (shown in Figure 12(d)). The system can detect EVs as low as s $10^7$ EVs/ml within 2 minutes based on the induced structural color change upon capture of the EVs on the metasurface.

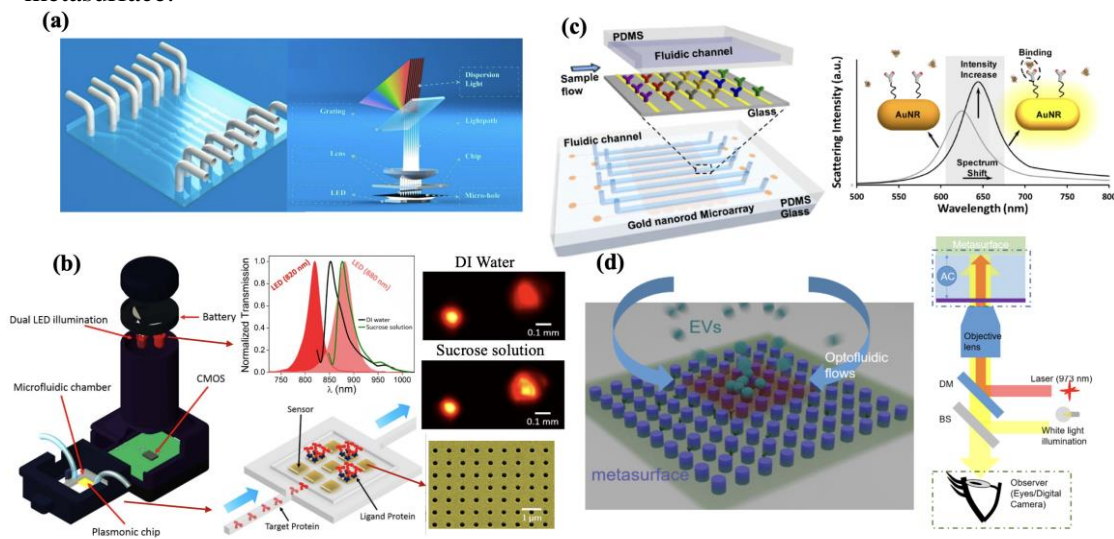

**Figure 12** (a) The top left panel shows schematic of the microfluidic chip. The right panel panel depicts the working principle of the smartphone biosensor system with the multi-testing-unit. (a) adapted with permission from ref. [189]. Copyright 2020 Multidisciplinary Digital Publishing Institute. (b) The left panel shows the schematics of the on-chip optofluidic sensor. The inset shows the microfluidic chamber and the transmission spectral of dual LED illumination and the resonance shift for different immerse solution. The right panel shows the COMS images under different conditions and the SEM image of gold nanohole array. (b) adapted with permission from ref. [190]. Copyright 2014 Springer Nature. (c) presents the schematic of the LSPR microarray chip. Nanorod microarrays were integrated in a microfluidic chip with eight parallel microfluidic detection channels consisting of inlet and outlet ports for reagent loading and washing. The left figure shows that binding of the analyte molecules to the receptors induces a redshift of spectrum. (c) adapted with permission from ref. [193]. Copyright 2015 American Chemical Society. (d) shows the experimental set-up for generating ETP flow and detecting the color change of metasurface. When EVs binds to the surface, the refractive index change will cause color change. (d) adapted with permission from ref. [104]. Copyright 2024 Optica Publishing Group.

### 3.1.2 Bound states in the continuum-assisted optofluidic sensors

Bound states in the continuums (BICs) have emerged as a hot topic in optofluidic sensor designs due to its high Q factor and strong light matter interaction over other dielectric modes such as Mie resonances and Fano-like modes. Moreover, the BIC modes in dielectric structures do not have the severe photothermal heating present in plasmonic structures. This makes it compatible with biological applications. Recently, optofluidic sensors based on BICs are presented with excellent performance [188]. For example, Romano *et al.* [194] have employed a BIC metasurface with two distinct resonance peaks, achieving considerable sensitivity of 178 nm/RIU to changes in refractive index. A microfluidic chamber (shown in Figure 13(a)) was integrated into the device to infiltrate different refractive index liquids while monitoring in real-time the resulting position of the resonance wavelength. In another work from Romanno *et al.* [195], they make use of the BIC metasurface and microfluidic chamber to get refractive index hyperspectral images. The microfluidic channel helps them conduct time-resolved monitoring of the refractive index changes. In other to boost the sensitivity of optofluidic BIC sensors, Liu *et al.* [196] have developed BIC metasurface (shown in Figure 13(c)) with a high-Q factor of up to 1200, achieving a sensitivity of $2.7 \times 10^4$ deg/RIU when the refractive index changes. They also integrated their BIC structure inside a microfluidic chamber to make the sensor measurements faster, easier, and more reproducible. Hyperspectral imaging-based optofluidic BIC sensors also hold promise, as they enable label-free detection without the need for a spectrometer. As shown in Figure 13(b), Jahani *et al.* [14] have crafted dimer BIC metasurface sensor units within a microfluidic chip. The chip consists of three independent channels and each channel has 3 sensor units. By utilizing the flow channels, their optofluidic sensor has enabled an average detection limit of 0.41 nanoparticle/μm² and real-time quantification of EVs binding from solutions as low as the equivalent of 204 femtomolar.

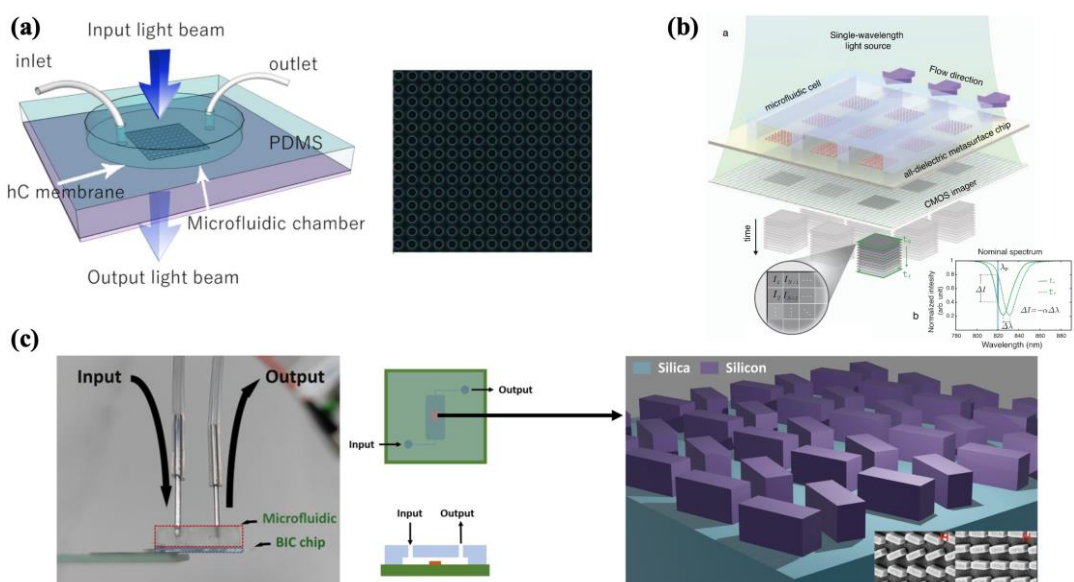

**Figure 13** (a) Schematic of the optofluidic set-up for the work by Romano *et al*. The microfluidic chamber was made by PDMS and the inlet and outlet control the fluid. The right panel shows the SEM image of the nanophotonic structure, which consists of air holes arranged in a square lattice. (a) adapted with permission from ref. [194]. Copyright 2018 Chinese Laser Press. (b) Sketch of the optofluidic set-up for the hyperspectral imaging-based optofluidic BIC sensor. The 2D microarray of all-dielectric sensors integrated with a microfluidic cell consists of three independent flow channels. The insight shows that the transmission intensity will change when analytes are bonded to the surface. (b) adapted with permission from ref. [14]. Copyright 2021 Springer Nature. (c) Device photograph of the microfluidic BIC chip. With the assistance of pumps, the flow can get in(out) at the input(output) end. The right panel shows the schematic of BIC metasurface and SEM image of fabricated samples. (c) adapted with permission from ref. [196]. Copyright 2023 American Chemistry society.

### 3.2 Optofluidic surface-enhanced vibrational spectroscopies

The optofluidic surface enhanced vibrational spectroscopies (such as SEIRA and SERS) sensors are rapidly becoming a vital tool for the in-situ detection of particles such as exosomes and proteins. Considering that water has large absorption at IR region, biochip materials and thicknesses need to be carefully selected [197,198]. As shown in Figure 14(a), Kavungal *et al.* [199] designed SEIRA gold nanorod array combined with microfluidics to facilitate capture and analysis of target protein in minute sample volumes. The home-made flow-cell helps to continuously run the buffer and inject samples over the sensor and retrieve time-resolved spectral signals every several seconds.. The designed micro-flowcell with less than 35 nL channel volume helps to reduce the amount of required samples and amplifies the antibody-antigen interaction, thereby leading to excellent binding performance. Moreover, the microfluidic chip consists of three independent microfluidic channels that can capture different proteins simultaneously. Furthermore, Xu *et. al.* [200] has developed an $Al_2O_3$-based hybrid plasmonic-nanofluidic SEIRA sensor that precisely guides analytes to hotspot regions via microchannels (shown in Figure 14(b)). This microfluidic approach significantly improves the utilization of enhanced electromagnetic field and makes it possible to distinguish a small concentration change of 0.29% for acetone molecules. Fu *et al.* [201] presented an optofluidic SERS measurement of 3,3',4,4'-tetracholorbiheny (PCB77), as shown in Figure 14(c). With the assistance of microfluidic technology, they made the aqueous environment closer to the biological environment in human samples and achieved a low detectable concentration as low as $1.0 \times 10^{-8}$ M. Benefiting from laser printing to construct nanogap-enabled SERS structures in microchannels, Lao *et al.* [202] presented a microfluidic SERS sensor with the maximum enhancement factor of SERS signal ~ $8 \times 10^7$. Compared with conventional SERS without dynamic flow and environmental control, the microfluidic SERS device provides real-time monitoring, ease of integration, and reusability but also higher reproducibility profiting from effectively dissipating heat. Jalali *et al.* [203] demonstrated a microfluidic nano-structured devices that employs SERS for the identification of EVs. They experimentally show that fluidic channels can generate monolayer distribution of EVs and dramatically increase the probability of EV existence in the laser illumination area, which enable distinguishing EV subpopulations from two glioblastoma cell lines.

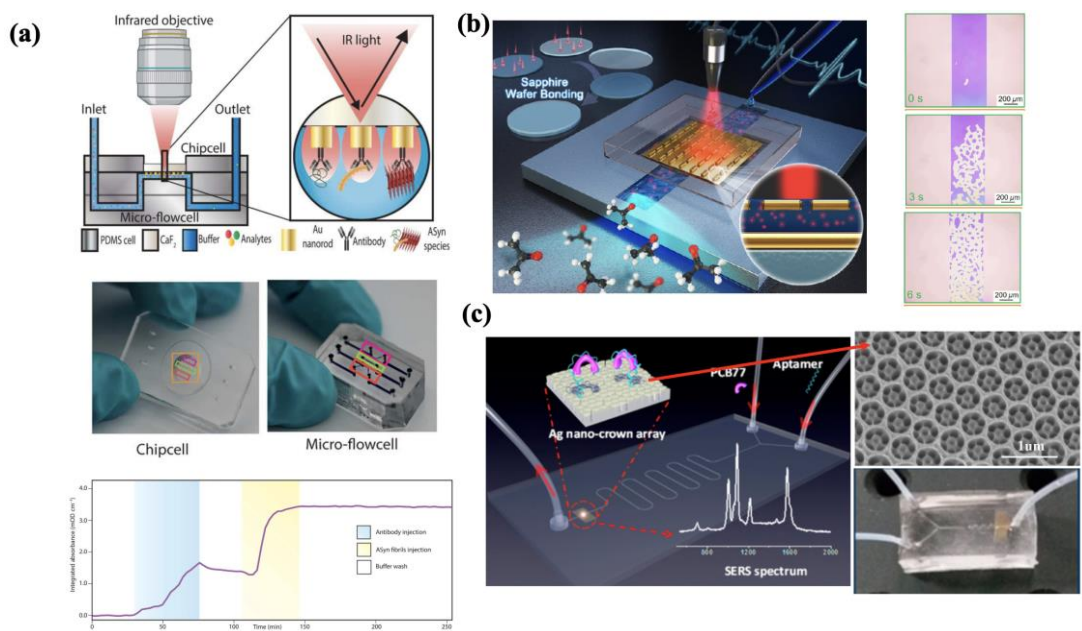

**Figure 14** (a) shows the schematic of the optofluidic setup used for backside-reflected SEIRA measurements. The Insight shows capture of all aSyn structural species using antibodies. The middle panel shows the fabricated chip cell and the micro flowcell follows the three-row design of the sensor. The bottom panel shows the time resolved reflectance measurements. (a) adapted with permission from ref. [199]. Copyright 2023 AAAS. (b) presents the concept of the self-driven 3D plasmonic $Al_2O_3$-based mid-infrared SEIRA sensing platform. The left panel shows that the liquid analytes can be delivered to sensor region through capillary force without any external actuation. (b) adapted with permission from ref. [200]. Copyright 2020 American Chemical Society. (c) The left panel shows the schematic of the aptamer-based SERS sensor. There are two inlets for injecting aptamer and PCB77 separately. The right panel shows the SEM of Ag nano-crown array and photograph of the device. (c) adapted from ref. [201]. Copyright 2015 American Chemical Society.

### 3.3 Optofluidics-enabled fluorescence sensors

Fluorescence-based sensing approaches are widely used for the quantitative detection of various labeled molecules. Microfluidic technology can significantly enhance the performance and capabilities of fluorescence sensors as it helps control the assay environment, reduce the reaction time, and offer high-throughput analysis. For example, Liu *et al.* [204] presented a fluidic thermophoretic sensor, which makes use of positive thermophoretic forces and buoyancy-driven fluidic flows to rapidly concentrate aptamer-bounded EVs for producing an amplified fluorescence signal through accumulation. They experimentally showed that this fluidic approach can enrich EVs by more than 1000 folds relative to free diffusion. The EV accumulation takes approximately 15 minutes after which the signal was analyzed to quantify the tumor-associated EVs. Similarly, Deng *et al.* [205,206] achieved the accumulation of viral particles through diffusiophoretic force generated by an infrared laser and a PEG-containing medium. Shown in the bottom of Figure 15(a), a 2100-flold enrichment of viral particles was rapidly accumulated within 15 minutes and the fluorescence intensity follows the similar trend (enhanced up to $8 \times 10^6$ photon counts per second). DNA computation is a powerful technology for analyzing molecular pattern of single bioparticle, but it is challenging as the tiny size of EVs and low amount of proteins in EV membranes. Li *et al.* [207] functionalized microbeads with biotinylated aptamers for capturing EVs after computation oprations. They utilized thermophoresis to accumulate the tumor-derived EVs for amplifying output fluorescence signal. Based on the DNA computation technology on EV membranes with significantly enhanced the signal, the proposed assay acceived a high accuracy of 97% for discrimination of breast cancer patients. As shown in Figure 15(b), Kang *et. al.* [208] demonstrated a mechanically flexible microfluidic fluorescence sensor integrated with silicon photodiode (Si-PD) and vertical cavity surface-emitting lasers (micro VCSELS). They successfully integrated the substrate with Si-PD and micro VCSELS in the elastomeric fluidic chips, achieving a high signal-to-noise ratio detection with multiplexed, real-time operation. Furthermore, nanostructures which enable fluorescence enhancement can also be integrated in microfluidic channels to achieve high sensitivity and compact design. In Figure 15(c), Iwanaga [209] integrated silicon nanorod structures in microfluidic channels, achieving an optical biosensor that can work in point of care settings. The microfluidic channels enable the quantitative control of several liquid reagents which is essential to make quantitative biosensors. With this combination of microfluidics and surface enhanced fluorescence structure, the platform can achieve detection at very small concentrations on the order of pg/mL. Tawa *et al.* [210] fabricated plasmonic grating in microfluidic channels, which achieves a 26.3 folds fluorescence enhancement and lower detection limit (100 pM). The four microfluidic channels in this design can flow either the same or different concentration solutions, which can significantly reduce the random error.

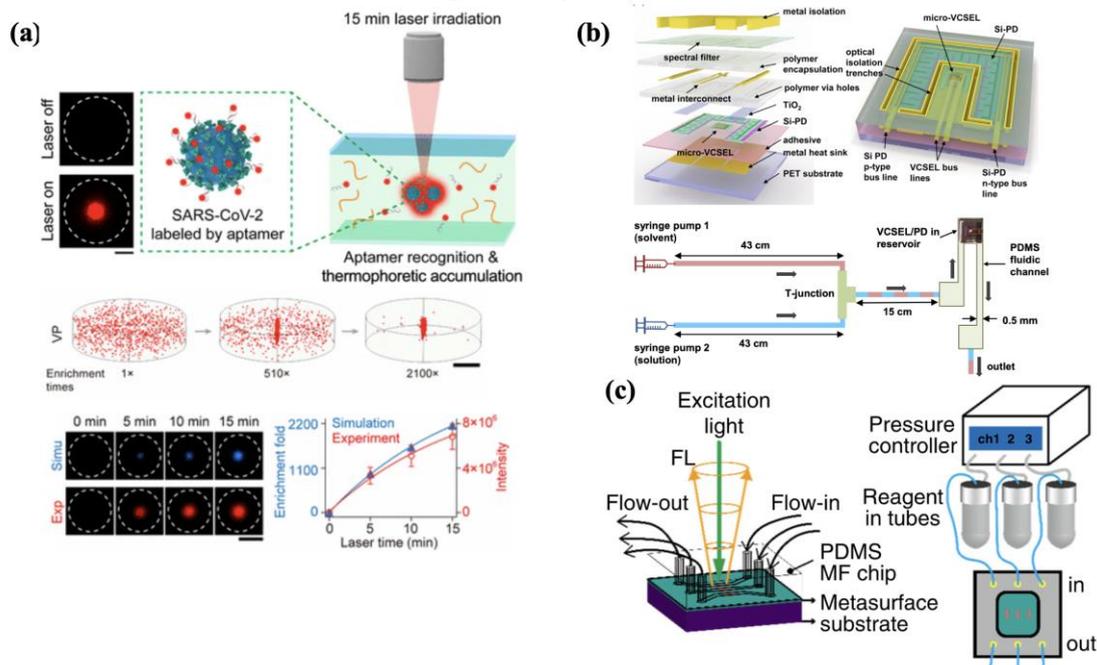

**Figure 15** (a) The top panel shows schematic of the one-step aptamer-based thermophoretic assay for rapid detection of SARS-CoV-2 viral particles. The middle and bottom panels show the numerical simulation and experimental results of accumulated viral particles upon 15 min of laser heating. (a) adapted with permission from ref. [205]. Copyright 2021 American Chemistry society. (b) The top panel shows the schematic design of mechanically flexible integrated fluorescence sensors based on heterogeneously integrated micro-VCSELs and silicon photodiodes (Si-PDs). The bottom panel shows depictions of the experimental setups for opto-fluidic measurements. (b) adapted with permission from ref. [208]. Copyright 2016 American Chemistry society. (c) Schematic illustration of the metasurface sensor chip and Schematic of the microfluidic setup, where the metasurface sensor chip is loaded in a holder. With the pressure controller, the flow rate can be precisely controlled ranging from 5-12 μL/min. (c) adapted with permission from ref. [209]. Copyright 2020 American Chemical Society.

# 4. Conclusion and Outlook

This review has presented a summary of latest advances and recently introduced concepts in optofluidics. There are many emerging opportunities both in fundamental biomedical science and translational applications where these recent developments are of immense importance. For example, optofluidic particle manipulation may be utilized to trap and perform single molecule analysis such as FRET, or Raman spectroscopy of single EVs to understand their heterogeneity, which is a subject of significant interest for which most of the tools available to biologists are currently unable to address.

On the translational side, optofluidics may be harnessed to rapidly capture and improve detection of cancer -associated markers and even for neurodegenerative conditions such as Alzheimer's. This review highlighted some recent works in this area, and we note that more opportunities exists to combine optofluidics manipulation with emerging sensing modalities such as CRISPR.

Apart from biological applications, the integration of Nanophotonics with Optofluidics may impact the actively investigated field of quantum information science. In quantum photonics, there is a big need to efficiently couple quantum emitters to nanophotonic cavities and circuits to enhance their emission properties via the Purcell effect. Traditionally these has been pursued via time-consuming and tedious process of AFM manipulation , multi-step lithography. We posit that Optofluidic concepts highlighted in this review represent a new paradigm for addressing this need with high throughput and represents an unexplored area of opportunity with grand potential.

**ACKNOWLEDGMENT**

Authors acknowledge support from National Institute of General Medical Sciences of the National Institutes of Health under award number R35GM150572.